# Thermal and Impact History of the H Chondrite Parent Asteroid during Metamorphism: Constraints from Metallic Fe-Ni

Edward R. D. Scott[1*], Tatiana V. Krot[1], Joseph I. Goldstein[2], and Shigeru Wakita[1,3]

[1]Hawai'i Institute of Geophysics and Planetology, University of Hawai'i at Mānoa, Honolulu, HI, 96822, USA

[2]Dept. of Mechanical and Industrial Engineering, University of Massachusetts, Amherst, MA 01003, USA

[3]Center for Computational Astrophysics, National Astronomical Observatory of Japan, 2-21-1 Osawa, Mitaka, Tokyo 181-8588, Japan

**Abstract:**

We have studied cloudy taenite, metallographic cooling rates, and shock effects in 30 H3-6 chondrites to elucidate the thermal and early impact history of the H chondrite parent body. We focused on H chondrites with old Ar-Ar ages (>4.4 Gyr) and unshocked and mildly shocked H chondrites, as strongly shocked chondrites with such old ages are very rare. Cooling rates for most H chondrites at 500 °C are 10-50 °C/Myr and do not decrease systematically with increasing petrologic type as predicted by the onion-shell model in which types 3 to 5 are arranged in concentric layers around a type 6 core. Some type 4 chondrites cooled slower than some type 6 chondrites and type 3 chondrites did not cool faster than other types, contrary to the onion-shell model. Cloudy taenite particle sizes, which range from 40 to 120 nm, are inversely correlated with metallographic cooling rates and show that the latter were not compromised by shock heating. The three H4 chondrites that were used to develop the onion-shell model, Ste. Marguerite, Beaver Creek, and Forest Vale, cooled through 500 °C at ≥5000 °C/Myr. Our thermal modeling shows that these rates are 50 × higher than could be achieved in a body that was heated by [26]Al and cooled without disturbance by impact. Published Ar-Ar ages do not decrease systematically with increasing petrologic type but do correlate inversely with cloudy taenite particle size suggesting that impact mixing decreased during metamorphism. Metal and silicate compositions in regolith breccias show that impacts mixed material after metamorphism without causing significant heating. Impacts during metamorphism created Portales Valley and two other H6 chondrites with large metallic veins, excavated the fast-cooled H4 chondrites around 3-4 Myr after accretion, and mixed petrologic types. Metallographic data do not require catastrophic disruption by impact during cooling.

# 1. INTRODUCTION

Ordinary chondrites contain information about the formation and thermal and impact histories of S-type asteroids as well as important constraints on the accretional history of planetesimals in the inner solar system and the evolution of the asteroid belt (e.g., Marchi et al., 2013). Here we focus on what can be learned about the metamorphic and early impact histories of the H chondrites from metallic Fe,Ni grains in relatively unshocked H3-6 chondrites.

If the parent bodies of the ordinary chondrites had been heated by [26]Al and allowed to cool undisturbed by impacts, they would have developed a so-called "onion-shell structure" with the most metamorphosed type 6 material occupying the central region surrounded by successive shells of less-metamorphosed type 5 through 3 material. Pellas and Storzer (1981), who introduced the term, found that cooling rates for six H4-6 chondrites derived from Pu fission track thermometry decreased with increasing petrologic type. They concluded that the parent body of the H chondrites had been heated by [26]Al decay and had cooled with an onion-shell structure.

Cooling rates of over 30 ordinary chondrites have also been inferred from Ni concentrations at the centers of zoned taenite grains (Wood, 1967; Willis and Goldstein, 1981b, 1983). Scott and Rajan (1981) and Taylor et al. (1987) found no evidence that the parent bodies had cooled with an onion-shell structure as cooling rates did not decrease systematically from type 3 to type 6. They concluded that either the parent bodies never had onion-shell structures or their interiors were rearranged by impacts, possibly by catastrophic disruption and reassembly, as Grimm (1985) suggested, prior to cooling below ~600 °C when taenite grains start to develop Ni concentration gradients. Because metallographic cooling rates for the six H chondrites studied by Pellas and Storzer (1981) were consistent with fission track cooling rates, Taylor et al. (1987) suggested that the apparent conflict between the two techniques was merely a result of the small sample size of the fission track data set.

The most detailed evidence for the onion-shell model was presented by Trieloff et al. (2003), who measured Ar-Ar ages and fission track cooling rates for nine H chondrites and compared them with the Pb-Pb ages of Göpel et al. (1994). Both sets of radiometric ages were found to correlate inversely with petrologic type and closely matched ages calculated for various depths in a 100 km radius asteroid that was heated by [26]Al to 850 °C in its central regions and cooled without disturbance by impact to ~100 °C in ~150 Myr. Trieloff et al. (2003) did not discuss the conflicting metallographic cooling rates of Taylor et al. (1987) but suggested that the samples studied by these authors had been modified by shock or impact heating. Wood (2003) concluded that Trieloff et al. (2003) had resolved the controversy in favor of the onion-shell model.

Subsequent studies of the radiometric ages of H chondrites or thermal modeling of their parent body have largely endorsed the conclusions of Trieloff et al. (2003). Kleine et al. (2008) found that Hf-W ages of five H4-6 chondrites including four of those studied by Trieloff et al. (2003) decreased with increasing petrologic type. They concluded that the parent body had cooled with an onion-shell structure after heating by [26]Al. Harrison and Grimm (2010) used radiometric ages and metallographic cooling rates for H chondrites to constrain various thermal models and concluded that nearly all data could be accounted for by an onion-shell body of radius 75-130 km. They questioned the accuracy and validity of some metallographic cooling rates and suggested that the relatively small number of non-conforming data might be attributed to impacts that left the parent body largely intact. Finally, Henke et al. (2012) and Monnereau et al. (2013) focused on the eight H chondrites for which precise radiometric ages are available at three different closure temperatures (Trieloff et al., 2003). They concluded that an onion-shell body with a radius 110-130 km that accreted rapidly (in <0.2 Myr) around 1.8-2.1 Myr after CAI formation could successfully reproduce their cooling histories without invoking any disturbance by impacts.

Constraints on the thermal histories of ordinary chondrites from silicates and oxides suggest that the parent bodies did not cool with an onion-shell structure. Cation ordering temperatures in orthopyroxene in eight H4-6 chondrites indicated similar cooling rates at ~400-500 °C (Folco et al., 1996). Olivine-chromite thermometry gave similar mean temperatures for H4, H5 and H6 chondrites of ~690-770 °C (Wlotzka, 2005). Kessel et al. (2007) also found several inconsistencies with the onion-shell model. Ganguly et al. (2013) measured compositional profiles across coexisting orthopyroxene-clinopyroxene, olivine-spinel, and orthopyroxene-spinel grains in five H4-6 chondrites and found all grains to be essentially homogeneous. From calculated compositional profiles, they inferred that the chondrites cooled very rapidly above 700 °C at rates of 25-100 °C/kyr, several orders of magnitude faster than the rates generally inferred from radiometric ages and metallography. Ganguly et al. (2013) concluded that the H chondrite parent body had been fragmented and then reaccreted prior to cooling below 700°C.

Substantial isotopic evidence suggests that [26]Al was the major source of heat for melting early-formed planetesimals and heating bodies that accreted later (e.g., Bizzarro et al., 2005; Kleine et al., 2005; Hevey and Sanders, 2006; Ghosh et al., 2006; Kruijer et al., 2012; Sanders and Scott, 2012). However, there is also evidence in chondrites for localized impact heating during metamorphism (Rubin, 1995, 2004). Impact heating of asteroids, even porous ones, is very inefficient (Keil et al., 1997). However, Davison et al. (2012) inferred that large projectiles impacting into highly porous chondritic targets could have buried enough impact melt at depth to cause localized slow cooling. In addition, impacts during metamorphism may have excavated hot chondritic rock so that it cooled more rapidly and mixed hot and cold material (Davison et al., 2013; Ciesla et al., 2013).

We have determined cooling rates for a large suite of H3-6 chondrites to elucidate the geological history of the H chondrite parent body during metamorphism. Specifically we evaluated the onion-shell model and clarified the possible role of impacts in heating and mixing material. Metallographic cooling rates at 450-600 °C were determined by measuring Ni concentrations at the centers of taenite grains, which became enriched in Ni during kamacite growth at 450-600 °C (Willis and Goldstein, 1981b, 1983; Taylor et al., 1987). Cloudy taenite was studied using the same techniques that were applied to iron and stony-iron meteorites to constrain relative cooling rates at 300 °C (Yang et al., 1997a; Yang et al. 2010a, b). We also evaluated shock effects in silicate (Stöffler et al., 1991) and metallic Fe,Ni (Buchwald, 1975; Bennett and McSween, 1996a) in the same suite of H4-6 chondrites to monitor their shock history.

To probe the earliest stages in the thermal and impact history of the H chondrite parent asteroid, we have focused on H chondrites with old Ar-Ar ages (>4.4 Gyr) and the least shocked samples (Stöffler et al., 1991). Chondrites with Ar-Ar ages of >4.4 Gyr are very largely unshocked or only mildly shocked (Trieloff et al., 2003; Bogard, 2011). Nearly all moderately and strongly shocked S4-S6 chondrites have Ar-Ar ages of <4.4 Gyr (Swindle et al., 2009) as well as altered Fe-Ni grains and pockets of silicate-troilite melt (Bennett and McSween, 1996a). We specifically included the H chondrite, Portales Valley, which clearly suffered an ancient impact, but is not heavily shocked (Ruzicka et al., 2005; Bogard and Garrison, 2009).

## 2. SAMPLES STUDIED

Table 1 lists the 30 chondrites that we studied and the sources and catalog numbers of the 53 thick and thin polished sections used to determine metallographic cooling rates, shock stages, and cloudy taenite dimensions. In addition to focusing on ancient, unshocked samples, we avoided breccias. Four chondrites in Table 1 are listed as brecciated by Grady (2000)—Bath, Forest City, Ochansk, and Sena. In addition, ALH 77299 (H3) contains an H4 clast (Scott, 1984). However, our sections of these five chondrites appear to be unbrecciated. All the chondrites are falls or—in the case of the four Antarctic chondrites, Landreth Draw and Estacado—fresh finds.

 The petrologic types listed in Table 1 are mostly from Van Schmus and Wood (1967); three are from Gomes and Keil (1980): Avanhandava, Conquista, and Marilia. Petrologic types for the remainder—Ankober, Ehole, Quenggouk, Ste. Marguerite, and the Antarctic samples—are taken from sources listed in the Meteorite Bulletin Database (see http://www.lpi.usra.edu/meteor/metbull.php). In only one case did we find any disagreement about the classification of chondrites. Trieloff et al. (2003) listed Sena as an H5 chondrite although Van Schmus and Wood (1967) classified it as a H4, and this is the classification given in meteorite catalogs and the Meteoritical Bulletin Database (http://www.lpi.usra.edu/meteor/metbull.php). Christophe Michel-Levy (1981) listed Sena as an H5 chondrite but it is not clear why as polysynthetically twinned low-Ca clinopyroxene is abundant in Sena, although virtually absent in type 5 chondrites (Huss et al., 2006).

Because the assignment of petrologic type can be ambiguous in the case of a few ordinary chondrites, we focused on type 4 and 6 H chondrites, which are readily distinguished, and only included five H5 chondrites, even though they are by far the most abundant type. Table 1 includes a very high proportion of H4 chondrites (50%) including two Antarctic samples, as we discovered three H4 chondrites with exceptionally fast cooling rates, and we wanted to know if they were abnormal.

Subtype assignments of the H3.6 to H3.8 chondrites are from Sears et al. (2013). Classifications of these chondrites are less certain as high type 3 and low type 4 chondrites are not readily distinguished (Grossman et al., 2009). However, whether they are high type 3 or low type 4 is not critical for this study. We included Tieschitz, which some workers classify as H/L3, as we did not find many suitable H3 chondrites. Results for Tieschitz are not plotted in the figures although they strengthen our conclusions.

There are gaps in Table 1 for various reasons. Some are for type 5 chondrites, which were less critical for our study. Shock stages were not obtained for all chondrites: in some cases we lacked a suitable thin section, in others, the chondrites had already been studied by Stöffer et al. (1991, 1992). Metallographic cooling rates were not redetermined for all of the chondrites studied by Taylor et al. (1987) and Willis and Goldstein (1981b). In some cases, cloudy taenite was not studied as we lacked a suitable section that could be etched and repolished.

## 3. TECHNIQUES

### 3.1. Shock effects

Shock stages for the H chondrites were assigned using standard polished thin sections and transmitted light microscopy (Stöffler et al., 1991). Twenty or more of the largest olivines showing high birefringence colors were examined for sharp extinction (S1), undulatory extinction (>2° variation) (S2), or sets of parallel planar fractures (S3). The shock stage was assigned for the highest level shown by at least 25% of the grains. Chondrites showing mosaicized olivine (S4), planar deformation features in olivine (S5) or plagioclase grains that had been converted to maskelynite (S6) were not included in this study.

Features in metallic Fe,Ni and troilite due to shock and reheating of ordinary chondrites have been described by Bennett and McSween (1996a) and for iron meteorites by Buchwald (1975). We looked for the following features: disturbed M-shaped Ni profiles in taenite, recrystallized kamacite, polycrystalline troilite, and fizz-textured or dendritic shock melted metal-sulfide intergrowths. We also looked for disturbed or patchy cloudy rims on taenite grains like those in the atmospherically heated rims of iron meteorites (Buchwald, 1975). Unzoned plessite grains (or "zoneless plessite"), which consist of sub-micrometer sized tetrataenite particles and kamacite and commonly form in type 5 and 6 chondrites, were once attributed to shock heating. However, these particles formed during slow cooling of isolated monocrystalline taenite grains, which partly transformed to martensite that decomposed (Reisener and Goldstein, 2003; Goldstein and Michael, 2006). Polycrystalline metal grains, which are prevalent in type 3 and 4 chondrites due to limited grain coarsening during metamorphism, are not evidence for shock or rapid solidification, contrary to Hutchison et al. (1981)—see Willis and Goldstein (1981a).

### 3.2. Cloudy taenite studies

The average size of the high-Ni phase in the cloudy zone was measured in 22 H chondrites. A FEI Magellan field emission scanning electron microscope at the University of Massachusetts was employed. The meteorite samples were polished with diamond paste (6 and 3 μm) and finished with colloidal $SiO_2$. The samples were ultrasonically cleaned in distilled water after each polishing step and were etched with 1 or 2% nital (nitric acid in alcohol) for 15 to 90 sec. On etching, the low Ni honeycomb phase in the cloudy zone is attacked and the zone appears dark brown in reflected light (Fig. 1a). After suitable cloudy taenite rims had been located optically, a metallic coating was applied to the section and the cloudy taenite was imaged at 5 keV in the scanning electron microscope. The structure of the cloudy taenite was observed at high magnification, typically 80 to 200 kX (Fig. 1b), and the sizes of the high-Ni particles were measured directly from a micrograph. Particles next to the boundary with the tetrataenite rim were measured where the local Ni concentration is 40-42 wt.% Ni (Yang et al., 1997a), using the procedure developed by Goldstein et al. (2009a). High-Ni particles farther away from the boundary were not measured as the high-Ni particle size and Ni content of taenite decrease with increasing distance from the boundary (Yang et al., 1997a). Except for Landreth Draw and Estacado where we were limited by the size of the polished section, tetrataenite particle sizes were measured in three regions of cloudy taenite in each of 3-6 taenite grains with 10 particles measured in each region.

## 3.3. Metallographic cooling rates

Kamacite and taenite grains in polished unetched sections were analyzed using a JEOL JXA8500 electron microprobe analyzer at a voltage of 20 kV and a beam current of 20 or 25 nA. Two major siderophile elements were measured (Fe and Ni) and one minor element (Co), plus the elements Cr, Ca, S, and Si. Although typically undetectable in metal, these four elements provided a check for beam overlap with troilite grains and non-metallic inclusions, which are common in type 3 chondrites. Pure metals were used as standards for Ni and Co, troilite for Fe and S. Counting times for peak and background were 20 and 10 s for Fe and Ni, 40 and 20 s for Co. The counting times for the Si and Ca peaks were 30 and 20 s respectively.

Metallographic cooling rates were determined from the central Ni concentrations of taenite grains using the techniques of Wood (1967). Like Taylor et al. (1987), we used the cooling rate curves calculated by Willis and Goldstein (1981b). Taenite grains were selected using low and high resolution Ni X-ray maps to avoid etching and repolishing sections. Equant grains with uniform Ni-rich rims were selected for analysis. We avoided taenite grains with extensive wide tetrataenite rims adjacent to troilite as well as grains that contained ubiquitous troilite blebs and those with Ni-enriched subgrain boundaries (as in Fig. 6 of Reisener and Goldstein, 2003). We excluded unzoned ("zoneless") plessite grains (Reisener and Goldstein, 2003) and tetrataenite grains. A small number of grains with very high central Ni contents of >40% were excluded because calculated cooling rate curves on the Wood diagram do not match measured compositions for taenite grains with >45% Ni (Hopfe and Goldstein, 2001).

The cooling rate curves of Willis and Goldstein (1981b) were determined using the P-free Fe-Ni system as Fe-Ni metal grains in ordinary chondrites have very low P concentrations of <500 ppm, and phosphides are absent (Reed, 1969) except in type <3.5 chondrites (Rambaldi and Wasson, 1981), which we did not analyze. Ruzicka et al. (2005) reported phosphides (rhabdites) in Portales Valley but we were not able to locate any in our sections suggesting that the Fe-Ni system is still appropriate for this chondrite.

The cooling rate curves for ordinary chondrites have not changed since the work of Willis and Goldstein (1981b), unlike those for iron meteorites. This is because there are no complications for metal in chondrites due to the effects of P on diffusion rates and phase equilibria, as discussed above. In addition, taenite grains in chondrites have grain boundaries, which are virtually absent in iron meteorites and provide ready nucleation sites for kamacite growth in ordinary chondrites (Reisener and Goldstein, 2003). Thus, kamacite nucleation in ordinary chondrites does not require prior formation of phosphides or martensite as it does in iron meteorites (Goldstein et al., 2009b).

## 4. RESULTS

### 4.1. Shock effects

Table 2 shows the shock stages that we assigned to 23 H chondrites. Cases where the proportion of grains with the highest shock level was close to the 25% cutoff are indicated by parentheses (e.g., Avanahandava S1(2)). In the H3 chondrite, ALH77299, we found a mixture of chondrules and other materials that had been shocked to different shock stages: S1 to S3, consistent with its classification as a post-metamorphic fragmental breccia (Scott, 1984).

Table 2 also shows published shock stages for these and eight other chondrites. In most cases the shock classifications are the same or differ by one shock stage. The exception is Beaver Creek, which is S1 according to Stöffler et al. (1992) and our study, but S3 according to Rubin (1994). In cases of disagreement, we use our data or those of Stöffler et al. (1992). Eighteen of the 31 chondrites in Table 1 (58%) are shock stage S1, nine (29%) are S2, and four (13%) are S3. By comparison, Stöffler et al. (1992) found 30% S1, 20% S2 and 37% S3 in their suite of H chondrites. Thus our suite is strongly biased towards unshocked and low-shock samples, as we intended.

Shock or reheating effects in kamacite and taenite grains were notably absent except in Menow where there are some fizz-textured troilite-metal intergrowths, also called type 1 metal-troilite intergrowths, which form by highly localized shock melting in many types of meteorites (Buchwald, 1975; Scott et al., 2010; Yang et al., 2014). They are present in a few shock stage S1-S3 chondrites, but contrary to Bennett and McSween (1996a), they are very rarely found in higher shock stages. Ramdohr (1973) shows a picture of fizz-textured troilite-metal texture in Bath but it was not present in our section.

## 4.2. Cloudy taenite studies
Cloudy taenite was observed in all but four of the 26 chondrites that were etched and studied using scanning electron microscopy. The mean sizes of the high-Ni particles for the 22 chondrites range from 42 to 125 nm and are listed in Table 3 with the number of particles analyzed in each chondrite. We found small systematic differences between regions in some chondrites, possibly due to orientation effects, as the particles are not spherical. The error in the mean value for each chondrite was therefore calculated from the mean particle size in each region. Values for twice the standard error of the mean are listed in Table 3 (2 SEM) and range from 2-6 nm for most chondrites.

Figure 2a shows the mean cloudy taenite particle size and 2 SEM values for the 21 H chondrites arranged according to their petrologic type. (The H/L3 chondrite Tieschitz is not plotted.) Mean cloudy taenite particle size tends to increase with petrologic type but the correlation is weak as there is considerable overlap between the cloudy taenite particle sizes of different petrologic types. (The correlation coefficient is 0.40, which is significant at the 93% level, but without Dhajala the values are 0.26 and 74%.) Three H6 chondrites have especially coarse cloudy taenite with mean particle sizes of 100-125 nm: one H3 chondrite has especially fine-grained cloudy taenite with a mean particle size of 42 nm wide. The remaining chondrites have particle sizes of 60-100 nm. Because of the precision of the data, relative cooling rates are well constrained. For example, the H4 chondrites, Ankober and Ochansk, have larger taenite particle sizes and cooled slower at 300 °C than the H6 chondrites, Mt. Browne and Butsura.

Figure 2b shows the same plot with symbols showing the shock stage of the chondrites, which range from stage S1 to S3. Cloudy taenite particle sizes show little systematic variation with shock stage: particle sizes for shock stage S2 and S3 chondrites are not systematically different from those for shock stage S1 chondrites. The three shock stage S2-S3 H6 chondrites have smaller cloudy taenite particles than the four S1 H6 chondrites, but we do not believe this is significant as H5 chondrites show the opposite trend. Two shock stage S3 chondrites, Mt. Browne and Ochansk, and the S(2)3 chondrite, Queens Mercy, have normal cloudy taenite: no effects of shock heating were observed. The three chondrites lacking cloudy taenite are discussed in § 5.1.4.

Cloudy taenite particles in the chondritic portion of Portales Valley and in a metallic vein with Widmanstätten pattern are indistinguishable in size, 109±5 and 106±7 nm respectively, as Sepp et al. (2001) first reported. Their values of 123±4 nm for vein metal and 117±3 nm for the chondritic portion have similar precision but are 12% higher. This difference can easily be explained by variations in the measurement procedure rather than any thermal or chemical gradient in Portales Valley.

### 4.3. Metallographic cooling rates

The taenite grains with uniformly wide Ni-rich rims that we selected have M-shaped Ni profiles (Fig. 3a shows a typical example). Figure 4a shows the central Ni concentrations in the selected taenite grains vs. the apparent distance to the nearest grain boundary for two H chondrites that represent most of the range of cooling rates observed: Kernouvé and Dhajala. The cooling rate curves for 1-100 °C/Myr are from Willis and Goldstein (1981b) and that for 1000 °C/Myr from Taylor et al. (1987) who extrapolated the 0.1-100 °C/Myr curves of Willis and Goldstein (1981b). Since the scatter is mostly caused by off-center sectioning of taenite grains, we follow these authors in drawing a curve through the lower envelope of the data points, making a small allowance for the errors in the Ni concentration (~2% relative) and distance (±1 μm). Interpolating between the curves assuming their spacing depends linearly on the log of the cooling rate gives values of 12 °C/Myr for Kernouvé and 85 °C/Myr for Dhajala, which we round off to 10 and 100 °C/Myr in Table 3. Taenite compositions are shown in Fig. 4b for Ankober (H4) and Butsura (H6): in this case, the lower petrologic type has a slower cooling rate, 7 vs. 50 °C/Myr. Metallographic cooling rates for 22 H3-6 chondrites were determined with this methodology using 9-28 taenite grains per chondrite (Table 3). The data are given in Table S1 in the Supplementary materials.

Central Ni concentrations of taenite grains ranged from 25-40% Ni except in the two fast-cooled H4 chondrites, Forest Vale and Ste. Marguerite, and in the largest grains in Dhajala and Menow. Taenite grains in Forest Vale, like those in Beaver Creek (Taylor et al., 1987; Fig. 1), have central Ni concentrations of 10-20% and steep-sided Ni profiles. All grains larger than 20 μm in radius in these two chondrites have central Ni contents of ~10%—the approximate bulk Ni concentration in metallic Fe-Ni in H chondrites—as cooling was too rapid for Ni to diffuse into the center of these grains after kamacite nucleated. The lower envelope of data points on the Wood plot for grains in Forest Vale with radii of 8-20 μm matched the 10,000 °C/Myr cooling curve in Taylor et al. (1987). This cooling rate curve, which is not shown in Fig. 4, was obtained by extrapolating the curves of Willis and Goldstein (1981b) for cooling rates of 0.1-100 °C/Myr.

Ste. Marguerite appears to have cooled faster than Beaver Creek and Forest Vale as the Ni profiles in the taenite rims are steeper: only the outermost few micrometers are enriched in Ni (Fig. 3b). Six grains 10-50 μm in radius have a mean central Ni content of ~10%. Since the four grains with radii of 10-20 μm plot below the $10^4$ °C/Myr curve, we infer that this is a lower limit for the cooling rate of Ste. Marguerite. We found no evidence for shock or reheating in Ste. Marguerite and Forest Vale and infer that they cooled rapidly after metamorphism.

In Menow, the larger taenite grains have central Ni concentrations below 25% and the M-shaped Ni profiles show fluctuations of up to ±5% Ni in the central regions. Etching shows that large grains have cores of martensite and the Ni fluctuations indicate decomposition of martensite into plessite at the sub-micrometer scale (Hutchison et al., 1981).

In Fig. 5a we have plotted metallographic cooling rates as a function of petrologic type for 22 H chondrites from our studies plus data from Taylor et al. (1987) and Willis and Goldstein (1981b) for 13 chondrites that we did not analyze (Tables 3 and 4). For Sutton, which was analyzed by both groups but not by us, we plot the geometric mean. Aside from the three fast cooled H4 chondrites, which were discussed above, nearly all the H chondrites have cooling rates between 5 and 50 °C/Myr with relatively minor differences between the petrologic types. Even if we exclude the three fast-cooled H4s, H4s have faster mean cooling rates than types 5 and 6, but the 1σ error bars of the means overlap. There are too few H3s in Fig. 5 to define a mean with any precision but two of the three are very close to the type 5-6 mean.

The shock stages of these chondrites are shown in Fig. 5b and do not correlate with cooling rate. The S2 chondrites and S3 chondrites appear to be randomly dispersed among the S1 chondrites. If we exclude the S2-S3 chondrites, cooling rates for petrologic types 4-6 show significant overlap in Fig. 5b. For the S1 chondrites excluding the fast-cooled H4s, there is a significant inverse correlation between type and log cooling rate but this depends heavily on Dhajala.

Given the scatter on the Wood diagram (Fig. 4), we estimate that the precision of our cooling rates in Table 3 is about a factor of 2, somewhat higher than the value of 1.5 given by Wood (1967) and Taylor et al. (1987). However, for the fast cooled H4, Forest Vale, and for Menow (see above), we note in Table 3 that the precision is double the normal value. In Table 4 we compare our metallographic cooling rates for seven chondrites with published cooling rates by Willis and Goldstein (1981b, 1983) and Taylor et al. (1987). Most values are within a factor of 2 but cooling rates for Guareña and Quenggouk differ by a factor of 3.

The accuracy of the cooling rates depends also on uncertainties in the diffusion rates and equilibrium phase compositions and was estimated by Wood (1967) to be a factor of 2.5. A simple comparison with cooling rates determined from the difference in the Pb-Pb and Ar-Ar ages and their estimated closure temperatures of 720±50 and 550±20 K (see § 5.2) using the best constrained cooling rates of Richardton, Kernouvé, and Guareña shows that the metallographic cooling rates are 1.4-4 × faster than those determined from radiometric ages. Uncertainties in the cooling intervals for Ste. Marguerite, Forest Vale and Allegan are too large for useful comparisons. Given the extrapolations from the 0.1-100 °C/Myr cooling curves of Willis and Goldstein (1981b), the cooling rate of $10^4$ °C/Myr listed in Table 3 for Forest Vale should only be taken as an order-of-magnitude estimate.

# 5. DISCUSSION

Increasing evidence suggests that [26]Al was the major heat source for causing melting and metamorphism of meteorite parent bodies, and that heating by electromagnetic induction, impacts, and the decay of [60]Fe were of minor importance (Ghosh et al., 2006). Bodies that accreted less than ~2 Myr after CAI formation were melted by [26]Al to form metallic cores and silicate mantles (Bizzarro et al., 2005; Kleine et al., 2005; Hevey and Sanders, 2006; Kruijer et al., 2012). Radiometric ages of chondrules suggest that the parent bodies of chondrites accreted later when [26]Al was abundant enough to cause metamorphism but not melting (Kita and Ushikubo, 2012; Sanders and Scott, 2012). Chondritic bodies that appear to have accreted 3-4 Myr after CAI formation were only mildly metamorphosed or aqueously altered (Fujiya et al., 2012, 2013; Sugiura and Fujiya, 2014). Although the assumption that [26]Al was homogeneously distributed has been challenged (Larsen et al., 2011), the Hf-W and Al-Mg ages of angrites, for example, now appear to favor a relatively homogeneous distribution in the inner solar system (Kruijer et al., 2014).

Radiometric ages and thermal models for H chondrites using [26]Al as the heat source are broadly consistent with this chronology and suggest that the parent body was 100-200 km in radius and accreted 2 Myr after CAI formation (Trieloff et al. 2003; Harrison and Grimm, 2010; Henke et al., 2012, 2013; Monnereau et al., 2013). However, the role of impacts during metamorphism of chondrites is still controversial. Did impacts provide significant heating, as Rubin (2004) argues, or were impacts so insignificant that the body cooled with an intact onion-shell structure as Trieloff et al. (2003), Henke et al. (2012, 2013), and Monnereau et al. (2013) have inferred? Did impacts excavate some type 6 material as the body cooled without greatly perturbing the onion-shell structure (Harrison and Grimm, 2010), or accelerate cooling due to the flow of heated materials to the surface (Ciesla et al., 2013), or were impacts during metamorphism large enough to cause disruption and reassembly of the H chondrite parent body (Taylor et al., 1987; Ganguly et al., 2013)?

To help answer these questions, we first address the validity of metallographic cooling rates of H chondrites (§ 5.1). Then we compare the radiometric ages and metallographic cooling rates of the set of chondrites studied by Trieloff et al. (2003) to see if they are consistent (§ 5.2). We then review Ar-Ar ages of H4-6 chondrites and compare them with metallographic constraints on their low-temperature cooling histories (§ 5.3). In § 5.4, we use thermal modeling to estimate the maximum cooling rate at 500°C in a body that was heated by [26]Al to see whether the fast cooled H4 chondrites could have been derived from a body that cooled with an undisturbed onion-shell structure. Next we briefly review models for impact heating of ordinary chondrites and constraints on the role of impacts in heating and mixing materials (§ 5.5). In § 5.6, we discuss evidence from microscopic features in ordinary chondrites and Ar-Ar ages of impact melted H chondrites and clasts for impacts during metamorphism. Finally, we discuss the impact and thermal history of the H chondrite parent body during metamorphism (§ 5.7).

## 5.1. Validity of metallographic cooling rates

Trieloff et al. (2003) did not discuss the metallographic cooling rates of Taylor et al. (1987), which were in conflict with the onion-shell model, except to say that "Our new results define the thermal history of the H chondrite parent body to a much greater precision than previous studies not using rigorously unshocked or undisturbed samples[15]". [Reference 15 was Taylor et al. (1987).] Clearly they believed that the cooling rates of Taylor et al. (1987) had been compromised by shock or impact heating after metamorphism. Whether they believed that all metallographic cooling rates were compromised or just some of them is unclear. However, since Trieloff et al. (2003), Henke et al. (2012, 2013), and Monnereau et al. (2013) did not make any use of metallographic cooling rates to constrain the thermal history of the H chondrite parent body, they may have concluded, like Lipschutz et al. (1989), that there could be inherent flaws in the metallographic cooling rates due to shock or impact heating. Since the validity of the metallographic cooling rates is crucial to many recent papers on iron and stony-iron meteorites (e.g., Yang et al. 2008, 2010a,b) as well as the conclusions in this paper, we first address whether metallographic cooling rates can be compromised by shock or impact heating. To do this we review the microstructures of Fe-Ni grains in slowly cooled meteorites, and then discuss how these features may be modified by shock and impact heating.

*5.1.1 Characteristics of metallic Fe-Ni in slow-cooled meteorites*

Residual taenite that remains after kamacite growth is unstable below 350 °C. Taenite with < 28% Ni can decompose via a shear transformation to form martensite (metastable distorted kamacite), which can then transform to kamacite and tetrataenite, ordered Fe,Ni with ~46-55% Ni (Reisener and Goldstein, 2003; Goldstein and Michael, 2006). Taenite with ~30-40% Ni decomposes on cooling below ~300°C via a spinodal mechanism to form what appears as a brown cloudy zone on etching. Separating the cloudy taenite from the kamacite, troilite, or silicate matrix is a rim of tetrataenite (Yang et al., 1997b; Reuter et al. 1988; Holland-Duffield et al., 1991; Uehara et al., 2011). In the scanning electron microscope, cloudy taenite appears as a sub-micrometer intergrowth (Fig. 1b) of rounded tetrataenite particles separated by thin walls of low-Ni martensite or kamacite. In iron and stony iron meteorites that cooled at between 5000 and 0.1 °C/Myr, the sizes of the cloudy taenite particles next to the tetrataenite rim, range from 10 to 500 nm in width and correlate inversely with the metallographic cooling rate (Yang et al., 1997a; Yang et al., 2010a, b; Goldstein et al., 2013). Thus the cloudy taenite particle size provides a valuable measure of the cooling rate at 300 °C, even though quantitative cooling rates cannot be derived directly using kinetic and thermodynamic data.

Cloudy taenite has been studied extensively in iron meteorites and pallasites in conjunction with studies of Ni zoning in taenite to constrain metallographic cooling rates at 300°C and understand the origin of these meteorites (Yang et al., 2008, 2010a, b; Goldstein et al., 2009a, 2009b). For mesosiderites, combined studies of cloudy taenite and Ni zoning in taenite were especially useful in unraveling the shock and thermal histories of mesosiderites and interpreting their Ar-Ar ages (Haack et al., 1996; Bogard, 2011). Cloudy taenite has also been studied in ordinary chondrites (Holland-Duffield et al., 1991), and its importance in evaluating relative cooling rates was recognized by Yang et al. (1997a).

Several other parameters in addition to cloudy taenite particle size are well correlated with metallographic cooling rates. These include the width of the tetrataenite rim at the kamacite-taenite interface (Goldstein et al., 2009a), and the concentrations of Ni in kamacite and taenite at their interface that are measured using electron microprobe analysis (Goldstein et al., 2014). Kamacite in equilibrium with taenite becomes progressively poorer in Ni on cooling below 400°C so that its Ni content at the kamacite-taenite interface is inversely proportional to the cooling rate.

*5.1.2 Characteristics of metallic Fe-Ni in impact-reheated meteorites*

All of the Fe-Ni phases found in slowly cooled meteorites are highly sensitive to shock deformation and impact reheating, and numerous effects can be distinguished according to the severity of the shock and the degree of reheating (e.g., Buchwald, 1975, Chapter 11; Bennett and McSween, 1996a). Kamacite that is shocked to pressures of 1-13 GPa contains Neumann twin lamellae and numerous dislocations. Above 13 GPa, kamacite is transformed to an ε phase with hexagonal structure, which reverts to distorted kamacite on pressure release giving a characteristic hatched appearance on etching. Mild post-shock heating causes strained kamacite to recrystallize.

Phases that form from residual taenite below 350 °C during slow cooling can be altered during mild reheating. Tetrataenite, produced by electron irradiation at 1 MeV, can be disordered at 600 °C in less than 30 minutes (Reuter, 1986). Calculations by Reuter (1986) indicate that tetrataenite will disorder in 3 hrs at 500 °C, 120 days at 400 °C and $10^3$ years at ~300 °C. The cloudy taenite intergrowth, which has dimensions of ~20 to ~600 nm, is especially sensitive to reheating, as can be observed near atmospherically heated surfaces of iron meteorites (Buchwald, 1975). Five IVA irons that were moderately or heavily shocked (13->40 GPa) lack cloudy taenite and show clear effects of reheating at the nanometer scale in the TEM (Goldstein et al. 2009b). In some cases, kamacite near the interface with taenite recrystallized and the interface was wavy. Grain boundaries formed allowing diffusion over distances of 150 nm or less from the interface so that the maximum Ni concentration measured with the TEM decreased to as low as 30% Ni. However, electron-microprobe measurements of the maximum Ni at the kamacite-taenite interface in these five reheated IVA irons using a nominal 1 μm wide x-ray source were not significantly different from the maximum Ni concentrations measured in unshocked irons with similar bulk Ni contents (Goldstein et al., 2009b, 2014). Cloudy taenite can therefore be removed by mild reheating before any change can be detected with the electron microprobe in the Ni concentration profile across taenite.

In meteorites that are more extensively heated by impacts, the steep Ni gradient in taenite at the interface with kamacite is smoothed out by diffusion and martensite decomposes to plessite, which may become spheroidized (e.g., Wood, 1967, p. 26; Axon et al., 1968; Buchwald, 1975, p. 134; Yang et al., 2011). In the most extensively reheated meteorites the original microstructures are entirely replaced by new intergrowths of kamacite, taenite and martensite, which are unlike the primary structures (Smith and Goldstein, 1977; Leroux et al., 2000; Yang et al., 2011). Laboratory simulations show that ordinary chondrites with such grossly modified microstructures cooled at rates of $10^{-2}$ to $10^5$ °C/yr (Smith and Goldstein, 1977). Strongly impact-heated chondrites have other characteristic features such as recrystallized or melted troilite and enhanced P concentrations in Fe-Ni phases (Smith and Goldstein, 1977; Bennett and McSween, 1996a).

There is a close connection between the level of shock recorded by the silicates and the degree of alteration due to post-shock heating recorded by the Fe-Ni (Bennett and McSween, 1996a). Thus shock stage S1-S2 chondrites lack evidence for reheating, except possibly for the rare presence of shock-melted troilite (see §5.1.3), whereas shock stage S4-S6 chondrites consistently show alteration of Fe-Ni phases to some degree. Because of variations in the post-shock environment and the different sensitivity of metal and silicate to reheating, the correlation between shock deformation in silicate and shock and thermal alteration of Fe-Ni is not exact. For example among shock stage S3 H chondrites, some like Ochansk, Marilia, and Mt. Browne have pristine Fe-Ni grains with cloudy taenite (Tables 2 and 3), whereas Orvinio was heated above 1000°C so that Fe-Ni grains were extensively melted and homogenized (Smith and Goldstein, 1977). Thus, the intensity of impact heating can be inferred more readily from metallic Fe-Ni grains than from silicate.

The presence of the sub-micrometer scale intergrowth in cloudy taenite guarantees that Ni concentrations at the centers of taenite grains have not been modified by diffusion of Ni through taenite over distances of many micrometers during shock or impact reheating. Chondrites with cloudy taenite therefore have metallographic cooling rates that were not compromised by shock heating. The possibility that major impacts could have reheated ordinary chondrites so that Fe-Ni grains were homogenized and then allowed to cool slowly again so that cloudy taenite could reform is discussed and excluded in §5.5.

*5.1.3. Cloudy taenite in H chondrites*
For the 22 H3-6 chondrites where we observed cloudy taenite (Table 3), we infer that the metallographic cooling rates of these H chondrites are very robust. Although the precise nature of the cloudy taenite intergrowth was not known when Wood (1967), Willis and Goldstein (1981b, 1983), and Taylor et al. (1987) determined metallographic cooling rates for ordinary chondrites, its presence has long been recognized as a characteristic of slowly cooled meteorites that have not been reheated (e.g., Buchwald, 1975). Because shock-heated ordinary chondrites are so common, chondrites were etched by Willis and Goldstein (1983) and Taylor et al. (1987) specifically to search for cloudy taenite before metallographic cooling rates were determined to eliminate reheated chondrites and also to allow taenite grains with symmetrical zoning to be selected. We are therefore confident that the metallographic cooling rates of these authors were not modified by shock reheating.

*5.1.4. H chondrites that lack cloudy taenite*
Cloudy taenite was not observed in three H4 chondrites, Forest Vale, Beaver Creek, and Ste. Marguerite, which have metallographic cooling rates of ~5000 to >10,000 °C/Myr and very steep Ni profiles near the edge of the taenite. [Beaver Creek was analyzed by Taylor et al. (1987)]. Judging from the dimensions of cloudy taenite dimensions in iron meteorites (Goldstein et al., 2013), we might have expected that the cloudy taenite particles in these meteorites would be around 10 nm in size. To better understand the lack of cloudy zone in fast cooled H chondrites, we prepared transmission electron microscope thin sections of Forest Vale and Ste. Marguerite with a focused ion beam instrument (Goldstein et al., 2009a). Selected kamacite/taenite interface regions were analyzed using a FEI Tecnai F30ST field emission transmission - analytical electron microscope at Sandia National Laboratories. Nickel X-ray scans at resolutions of 4 to 20 nanometers confirm that both chondrites lack cloudy taenite. The Ni enrichment at the kamacite-taenite interface that was measured with the electron microprobe analyzer results from a 100-400 nm wide, somewhat uneven tetrataenite band containing 50 to 55 wt% Ni. In addition, the taenite that was found to contain ~10-15% Ni with the electron microprobe contains nanometer-scale taenite or tetrataenite platelets with up to 50 wt.% Ni. We infer that Forest Vale and Ste. Marguerite had a two-stage cooling history. Cooling from ~700 to 400 °C was rapid enough to prevent the formation in taenite of a peripheral region with 30 to 42 wt% Ni, where cloudy taenite normally forms. Much slower cooling through ~350°C, or possibly reheating to this temperature, allowed tetrataenite precipitates to form in the central plessite and tetrataenite rims at the kamacite boundary, aided by grain boundary diffusion. Beaver Creek probably had a similar history.

The other H chondrite in which cloudy taenite was not observed in our study is Menow, which is a shock stage S1 chondrite (Table 2). Menow cooled at 60 °C/Myr, so it should have developed cloudy taenite. However, it is the only chondrite in Table 2 where we observed shock-melted metal-troilite intergrowths with fizz-textured textures. Photomicrographs of Menow in Hutchison et al. (1981) also show cloudy taenite borders with well-defined tetrataenite rims, implying that shock reheating was heterogeneous (Ashworth, 1981). Turner et al. (1978), who determined Ar-Ar ages for 16 relatively unshocked ordinary chondrites including many in our study (see § 5.3), found that Menow had lost more $^{40}$Ar (18%) than the others. They inferred that Menow had been reheated within the last 2.5 Gyr. This reheating likely caused the sub-micrometer decomposition of martensite into plessite in the cores of the taenite grains and the localized loss or degradation of cloudy taenite. However, apart from the Ni fluctuations in the cores of large grains, taenite still shows M-shaped Ni profiles with ≥45% Ni in the rim so the inferred cooling rate is probably reliable, albeit with a larger uncertainty because of the plessitic cores of large grains.

Some H chondrites that were shocked to S3 levels contain cloudy taenite (e.g., Ochansk and Mt. Browne). The lack of a definitive connection between the shock stage and the presence or absence of cloudy taenite probably reflects the sensitivity of the thermal effects to the initial conditions (such as temperature and porosity) and the volume of impact-heated material and post-shock cooling rate.

*5.1.5. Relationship between cloudy taenite particle size and cooling rate*
If the environment of the H chondrites was not altered significantly when they cooled from 550-450 °C to 300 °C, their metallographic cooling rates should correlate with their cloudy taenite particle size. Figure 6 shows an excellent inverse correlation between these parameters. In addition, the least squares line through the H chondrite data is very close to that determined for iron and stony-iron meteorites (Yang et al., 2010a, b; Goldstein et al., 2013). Thus the cloudy taenite particle sizes strengthen the argument that the metallographic cooling rates are robust.

The scatter for the H chondrites in Fig. 6 is larger than we might expect for quiescent cooling, given that the precision of the cooling rates is a factor of 2. It is possible that the outliers, Estacado, Guareña, and Portales Valley, which have very coarse cloudy taenite particles, may have experienced a significant decrease in cooling rate between 500 and 300 °C. Ganguly et al. (2013) inferred that Guareña had experienced such a decrease in cooling rate, but Kernouvé, which they linked with Guareña, plots on the least-squares line below Guareña.

*5.1.6. Cooling Rates of Type 3 Chondrites*
Two of the three H3 chondrites we analyzed cooled at 10-15 °C/Myr, slower than many H4-6 chondrites (Fig. 5). If type 3 chondrites had cooled in the outer layers of their parent asteroids, they should have cooled faster than all other chondrite types. Here we discuss whether cooling rates for type 3 chondrites could therefore be flawed, as Lipschutz et al. (1989) suggested.

One consequence of the low peak metamorphic temperatures of type 3 chondrites—around 600-675 °C (Huss et al., 2006; Harrison and Grimm, 2010)—is that type 3 chondrites may not have been converted entirely into taenite on heating, as is assumed for the cooling rate calculations. For H chondrites with a bulk metallic Fe-Ni composition of around 10 wt.% Ni, the minimum temperature for complete conversion to taenite is ~700 °C, according to the Fe-Ni phase diagram. Suppose an H3 chondrite had been heated to ~550 °C, for example, for long enough to form homogeneous grains of kamacite and taenite with a diverse range of sizes, then the compositions of its kamacite and taenite grains would be the same as those in a chondrite with a bulk Fe-Ni metal composition of around 25 wt.% Ni that had just entered the two-phase field on cooling. Willis and Goldstein (1981b) showed that the calculated curves on the Wood diagram are not sensitive to variations in the bulk Ni concentration in Fe-Ni in the range 7-30 wt.% Ni. Therefore the measured compositions at the centers of taenite grains in the H3 chondrite should define the same curve on the Wood diagram as taenite grains in an H6 chondrite that cooled at the same rate.

Cloudy taenite and cooling rate data for the three H3 chondrites plot very close to the line defined by the H4-6 chondrites (Fig. 6) confirming the validity of these cooling rates. We infer that cooling rates can be determined for type 3 chondrites as long as they were heated above ~550 °C for long enough to form homogeneous grains with an appropriate range of sizes. We note that some grains in the type 3 chondrites we studied could not be used for cooling rate analysis: e.g., taenite grains with numerous inclusions and those with complex, very fine-grained kamacite-tetrataenite intergrowths.

Although we determined metallographic cooling rates for only three H3 chondrites, there is additional evidence for slow cooling of type 3 chondrites. Tieschitz has a cloudy taenite particle size of 82 nm, which corresponds in Fig. 6 to a metallographic cooling rate of ~20 °C/Myr. It also has a whole-rock Ar-Ar age of 4.45±0.05 Gyr (Turner et al., 1978), within error of many other slowly cooled H chondrites (see §5.3.1). Göpel et al. (1994) note that two type 3 chondrites, Mezö-Madaras and Sharps, have similarly old Pb-Pb ages of 4.47-4.48 Gyr, and Scott and Rajan (1981) found that Mezö-Madaras has a metallographic cooling rate of 2 °C/Myr, allowing for the factor of two increase of Willis and Goldstein (1981b). We infer that many type 3 chondrites cooled at rates comparable to those of normal type 4-6 chondrites.

**5.2. Comparison with Trieloff et al. (2003) study**
If metallographic cooling rates are as robust as we claim, we should find substantial agreement between these cooling rates and constraints from radiometric ages. To test this, we list in Table 5 our data from Table 3 for the chondrites studied by Trieloff et al. (2003) with their Ar-Ar ages and the Pb-Pb phosphate ages of Göpel et al. (1994) expressed as the time between CAI formation and cooling to their respective closure temperatures of 550 and 720 K. [Nadiabondi (H5) is omitted from Table 5 as we lack metallographic data for this chondrite.] Note that Sena is listed in Table 5 as an H5 chondrite, which is the classification adopted by Trieloff et al. (2003), whereas in Tables 1-3 it is listed as an H4, as Van Schmus and Wood (1967) recommended.

Table 5 shows that the metallographic data for the Trieloff et al. suite are entirely consistent with their Pb-Pb and Ar-Ar ages. The three H6 chondrites, which have the longest cooling times to the Pb-Pb (45-63 Myr) and Ar-Ar closure temperatures (68-109 Myr), also have the slowest metallographic cooling rates (10-15 °C/Myr), and the largest cloudy taenite particle sizes (91-124 nm). The two H4 chondrites, which have the shortest cooling times for the Pb-Pb (4-6 Myr) and Ar-Ar chronometers (5-15 Myr), have the fastest metallographic cooling rates ($\geq 10^4$ °C/Myr), and cooled too fast to develop cloudy taenite. The three H5 chondrites (including Sena), which have intermediate cooling times for the Pb-Pb (11-17 Myr) and Ar-Ar chronometers (26-42 Myr), have intermediate metallographic cooling rates (15-20 °C/Myr) and cloudy taenite particle sizes (70-80 nm). If we had studied only the suite of chondrites studied by Trieloff et al. (2003), we would also have concluded that the H chondrite parent body had cooled with an undisturbed onion-shell structure.

Our comparison in Table 5 shows that metallographic cooling rates are entirely consistent with constraints from radiometric ages for the Trieloff et al. suite of H chondrites, which are very largely shock stage S1. In the previous section, we argued that our data for selected shock stage S2 and S3 ordinary chondrites should be considered as robust as those for the shock stage S1 chondrites. We therefore conclude that Trieloff et al. (2003) studied a group of H chondrites that were unrepresentative of those that escaped post-metamorphic impact heating.

Taylor et al. (1987) also compared metallographic cooling rates with the Pu fission-track cooling rates of Pellas and Storzer (1981), who favored the onion-shell model. Taylor et al. (1987) noted that for the chondrites analyzed by the two techniques, there was no inconsistency, and Lipschutz et al. (1989) reached a similar conclusion. Taylor et al. (1987) inferred that the six chondrites studied by Pellas and Storzer (1981) were unrepresentative.

There are two important differences between the suite of chondrites studied by Pellas and Storzer (1981) and Trieloff et al. (2003), which are listed in Table 6, and those studied by us and Taylor et al. (1987). In the former, the H6 chondrites have slower cooling rates than the H5 chondrites and all the H4 chondrites have exceptionally fast cooling rates. For the chondrites with metallographic cooling rates, there is considerable overlap between the cooling rates of H4, H5 and H6 chondrites, and the three fast-cooled H4 chondrites—Beaver Creek, Forest Vale, and Ste. Marguerite—appear to be aberrant. We studied 11 other H4 chondrites, seven of which are shock stage S1, and did not find any that cooled faster than 100 °C/Myr.

Other authors have also noted that many H4 chondrites did not cool rapidly, e.g., Benoit and Sears (1996) and Christophe Michel-Levy (1981). Table IV of Lipschutz et al. (1989), who compiled published and unpublished metallographic cooling rates, also identified relatively slowly cooled H4 chondrites. From the data in Tables 3 and 4 and the study of Benoit and Sears (1996), we estimate that ~90% of H4 chondrites with undisturbed Ni profiles in taenite cooled at rates comparable to those of H5 and H6 chondrites (22 out of 25). The set of chondrites studied by Trieloff et al. (2003) and Pellas and Storzer (1981), in which all H4 chondrites are fast-cooled, clearly is not an acceptable suite of typical, unshocked or low-shock H chondrites.

Unfortunately, many authors who determined radiometric ages for H chondrites have continued to focus on the suite of chondrites studied by Pellas and Storzer (1981) and Trieloff et al. (2003) without acknowledging that they are unrepresentative. As a result, the onion-shell model for H chondrites has been endorsed by many studies e.g., Göpel et al. (1994), Kleine et al. (2008), Henke et al. (2012), and Monnereau et al. (2013). We recommend that authors who focus on this suite of chondrites should clearly state that these chondrites are not representative of H chondrites with well-preserved metamorphic histories. They are invaluable for comparing different isotopic chronometers (e.g., Brazzle et al., 1999), but give a totally misleading view of the thermal and impact history of the H chondrite parent body.

### 5.3. Ar-Ar ages of H4-6 chondrites

#### 5.3.1. Age vs. petrologic type

Our conclusion that the systematic variation of Ar-Ar age with petrologic type found by Trieloff et al. (2003) is an artifact arising from an unrepresentative sample can be tested by examining the Ar-Ar ages of other unshocked or lightly shocked H chondrites. Figure 7 shows that five Ar-Ar ages of H4-6 chondrites from Turner et al. (1978), one by Bogard and Garrison (2009) and three by Schwarz et al. (2006) do not follow the same trend as the Trieloff et al. (2003) data (shown as red diamonds). Ar-Ar ages determined by Turner et al. (1978), Trieloff et al. (2003) and Schwarz et al. (2006) for the same chondrites are in close agreement showing that there is no systematic bias between these studies. Trieloff et al. (2003), who noted that the errors in the ages of Turner et al. are probably smaller than the quoted values, argued in their Supplementary Information that some chondrites studied by Turner et al. (1978) had lost Ar due to shock heating.

Four of the five H chondrites dated by Turner et al. (1978) and not by Trieloff et al. (2003) do have higher shock levels than those listed in Table 4. In the case of Ochansk, Trieloff et al. (2003, Suppl. Materials) suggest that the whole rock age of Turner et al. (1978) probably reflects some Ar loss from the matrix. However, the plateau ages of Turner et al. for Butsura, Mt. Browne, and Queens Mercy, which are also shock stage S2-S3 chondrites, are comparable in quality to their data for the S1 H6 chondrites, Kernouvé and Guareña. In any case, the H6 chondrites, Butsura, Queen's Mercy and Mt. Browne, have older Ar-Ar ages than the shock stage S1 H6 chondrites, so the difference cannot be attributed to impact heating, and the Ar-Ar age of Mt. Browne was confirmed by Schwarz et al. (2006). As noted in § 5.1.4, the H4 S1 chondrite, Menow, which was dated by Turner et al. (1978), did lose ~20% of its Ar due to impact heating, but it still has a fairly well developed high temperature plateau age. [Note that the extent of Ar loss is controlled by the post shock thermal history, not the peak shock level (Bogard, 2012).]

In an abstract that Trieloff coauthored, Schwarz et al. (2006) reported that Ar-Ar ages for four H4 chondrites range from 4526±6 to 4448±7 Myr. The former value matches the old ages of Forest Vale and Ste. Marguerite, but the latter value (which is plotted in Fig. 7) is lower than all but one of the H4-6 chondrites studied by Trieloff et al. (2003). Thus, there are three H4 chondrites that have Ar-Ar ages that match those of the H6 chondrites and one H5 chondrite, Monahans, with an Ar-Ar age of 4533±6 Myr (Bogard et al., 2001) that is as old as the two H4 chondrites analyzed by Trieloff et al. (2003). Published Ar-Ar ages therefore support our conclusion that the Trieloff et al. suite of chondrites is not representative and that Ar-Ar ages for additional chondrites are not compatible with undisturbed cooling of an onion-shell structure.

*5.3.2. Ar-Ar ages and metallographic data*
Despite the absence of any correlation between Ar-Ar age and petrologic type, Ar-Ar age and metallographic cooling rate are correlated for low-shock ordinary chondrites such that younger chondrites tend to have the slowest coolest rates (Haack et al., 1996). For H chondrites, this correlation relies entirely on the exceptionally fast cooled H4s, Forest Vale and Ste. Marguerite.

More impressive is the negative correlation between Ar-Ar age and cloudy taenite particle size, which does not depend on the two fast cooled chondrites (Fig. 8). Since the cloudy taenite particle size reflects the cooling rate around the Ar-Ar closure temperature of ~300 °C, we can infer that the time to cool to 300 °C was inversely related to the cooling rate at this temperature. The implications of this relationship are discussed in § 5.8 after we address the possible role of impacts in heating the parent bodies of ordinary chondrites.

**5.4. Thermal modeling of onion-shell body**
Three H4 chondrites have metallographic cooling rates of ≥5000 °C/Myr—Beaver Creek, Forest Vale, and Ste. Marguerite (Table 3). Rapid cooling at these rates from temperatures above 600 °C would not be expected in a body that was metamorphosed by heat from the decay of $^{26}$Al. If $^{26}$Al had been uniformly distributed within the body, temperatures would have increased almost isothermally during $^{26}$Al decay throughout the body except in the thin outer crust where heat could have been conducted to the surface on a timescale comparable to the half-life of $^{26}$Al. Thus material that was close enough to the surface to cool in less than 0.7 Myr, the half life of $^{26}$Al, would not have been heated significantly. Since cooling rates cannot be readily inferred from published thermal models for H chondrite parent bodies (e.g., Bennett and McSween, 1996b; Kleine et al., 2008), we have done additional thermal modeling.

Cooling rates at 500 °C were determined at various depths using a thermal model like that of Akridge et al. (1988) and Harrison and Grimm (2010) for bodies with radii of 50, 100, and 200 km, assuming instantaneous accretion at 1.5, 2.0, 2.4, and 2.8 Myr after CAI formation. [Preferred literature values are 50-130 km for the radius and accretion at 1.8 to 2.5 Myr.] The assumed initial $^{26}$Al/$^{27}$Al ratio in CAIs was $5 \times 10^{-5}$ and at these accretion times the $^{26}$Al/$^{27}$Al ratio would have been $1.1 \times 10^{-5}$, $7.6 \times 10^{-6}$, $5.0 \times 10^{-6}$ and $3.3 \times 10^{-6}$, respectively. Temperature-dependent values for specific heat and thermal conductivity were taken from Yomogida and Matsui (1983, 1984) and Harrison and Grimm (2010), assuming a fixed porosity of 6% (Akridge et al., 1988). The concentration of Al in the parent body was taken as 1.17 wt.% (Harrison and Grimm, 2010), and the ambient temperature was assumed to be 170 K. Full details of the thermal modeling are given by Wakita et al. (2014).

Figure 9 shows the cooling rate at 500 °C as a function of depth for bodies with radii of 50 and 100 km that accreted 2.0 Myr after CAIs. Peak cooling for both bodies occurs at a depth of about 4 km. At shallower depths the cooling rate at 500 °C rapidly decreases reaching zero at a depth of ~3 km where the peak temperature was exactly 500 °C. In practice, since metallographic cooling rates are measured over a range of temperature (e.g. 600-450 °C), we should expect that metallographic cooling rates would decrease monotonically with increasing depth. Figure 9 shows that peak cooling rates would have been around 80 °C/Myr for a body that accreted at 2.0 Myr after CAI formation, and would not have varied significantly with parent body size. Peak cooling rates increase as the time of accretion decreases. For a body that accreted earlier at 1.5 Myr, the peak cooling rate would have reached 100 °C/My, but peak temperatures would have exceeded 1400 °C causing extensive melting. Accretion at 2.4 Myr would not allow type 6 metamorphic temperatures to be achieved. Thus chondrites that cooled faster than 1000 °C/Myr at 500 °C could not have cooled in a body with an onion-shell structure, as Akridge et al. (1998) previously noted. The three fast-cooled H4 chondrites, Beaver Creek, Forest Vale, and Ste. Marguerite, which appeared to Trieloff et al. (2003), Pellas and Storzer (1981), Monnereau et al. (2013) and many other authors to provide key evidence in favor of the onion-shell body, could not have cooled in such a body. We infer that they were excavated by impact so that they cooled very rapidly from 700-400 °C, and then cooled more slowly through ~350°C beneath an ejecta blanket.

Plagioclase grains in the three rapidly cooled H4 chondrites contain excess $^{26}$Mg due to $^{26}$Al decay showing that they were cooled ~5-7 Myr after CAI formation (Zinner and Göpel, 2002; Telus et al., 2014). Ste. Marguerite, which has the highest inferred cooling rate, $>10^4$ °C/Myr, contains the highest inferred initial $^{26}$Al /$^{27}$Al ratio and Beaver Creek, which has the lowest cooling rate, 5000 °C/Myr, has the lowest initial $^{26}$Al /$^{27}$Al ratio (although within error of the Forest Vale value). Additional evidence for rapid cooling of Forest Vale was obtained by Ganguly et al. (2013) who derived a cooling rate of ~50,000 °C/Myr from 750°C by modeling the compositional zoning in adjacent orthopyroxene and spinel. Allowing for the errors, metal and silicate cooling rates are consistent. All three fast-cooled H4 chondrites may have been excavated by a single impact 5-6 Myr after CAI formation, about 3-4 Myr after the H chondrite parent body accreted.

Calculated cooling rates are shown in Fig. 10 as a function of peak temperature for a 100 km radius body that accreted at 2.0 Myr and was not subsequently impacted. Although peak temperatures in this model exceeded the FeS-Fe eutectic temperature of 988 °C, this is not a significant issue here as we may not have sampled materials from these regions. Alternatively, later accretion may have lowered the peak temperature below the eutectic temperature.

Figure 10 also shows the burial depths for various peak temperatures and very approximate temperatures for the boundaries at the type 3-4, 4-5, and 5-6 transitions (Harrison and Grimm, 2010). Neglecting the narrow zone near 3.3 km, we see that cooling rates should decrease systematically with increasing petrologic type. Even allowing for some misclassified chondrites near the boundaries and the precision of the cooling rates—a factor of ~2—this is contrary to Fig. 5 where the cooling rate ranges within types 3 to 6 are very similar, neglecting the three fast cooled H4 chondrites.

## 5.5. Role of impacts during and after metamorphism

A major weakness of simple onion-shell models like that described above is that they neglect the role of impacts in heating and mixing material. Although impacts cannot cause global heating and melting of asteroidal bodies, even porous ones (Keil et al., 1997; Davison et al., 2012), several authors have nevertheless proposed that impacts into highly porous targets were the dominant heat source for the metamorphism of ordinary chondrites and the formation of certain igneous meteorites (e.g., Wasson and Kallemeyn, 2002; Rubin, 2004). Rubin (2004) argued that all equilibrated ordinary chondrites had been shocked and locally melted then annealed to remove evidence of shock heating from S1 and S2 chondrites, However, deformation to shock stage S4 and above would generate mosaicized olivine in which the crystal lattice is severely distorted. During subsequent metamorphism, mosaicized chondule phenocrysts would be destroyed by recrystallization, contrary to what is observed in shock stage S1-3 chondrites. In addition, shock to stage S4 levels would only raise the temperature by 250-350 °C (Stöffler et al., 1991), which is insufficient for metamorphism of type 4-6 chondrites. Rubin (2004) invokes repeated impacts on surface material to cause metamorphism. However, on small asteroidal bodies, impact heated material tends to be lost and nearly all metamorphosed chondrites cooled slowly at depth at 500 °C, not rapidly near the surface.

Davison et al. (2012) have investigated whether impact heating of porous asteroids might have generated large volumes of impact melt that were deeply buried causing localized heating effects that mimic global metamorphism. They showed that high velocity impacts of a 25 km radius projectile into very porous asteroids could bury significant volumes of melted projectile at depth causing local heating to type 6 levels and cooling at rates of 1-100 °C/Myr. Below we discuss constraints on this impact heating model and the role of impacts in mixing material from impact melts and breccias among ordinary chondrites.

### 5.5.1 Impact melts

The impact heating model of Davison et al. (2012) predicts that we should find slowly cooled impact melted material. However, apart from Portales Valley, which is discussed in § 5.5.4, impact melted chondrites appear to have cooled rapidly as small bodies. H chondrite impact melts were quenched near the surface on timescales of minutes or hours, not cooled slowly at depth (e.g., Folco et al., 2004; Swindle et al., 2009; Wittmann et al., 2010). Impact melts in ordinary chondrites are glassy and fine-grained and the metal-sulfide textures indicate rapid cooling of small volumes of material (e.g., Bogard et al., 1995). If impact heating from deeply buried impact melts had been common, we should find significant volumes of chondritic impact melts with ages >4.4 Gyr, but they are rare among ordinary chondrites (see § 5.6). In addition, there should be slowly cooled chondrites with Ar-Ar ages that date the late heavy bombardment. Several impact melts did form in H chondrites at ~3.8 Gyr but they cooled rapidly (Folco et al., 2004; Swindle et al., 2009).

### 5.5.2. Regolith breccias

The effect of impacts on ordinary chondrites after metamorphism can be inferred from the properties of silicates and metal grains in regolith breccias. About 15% of H chondrites are regolith breccias as they contain solar-wind gases showing that some grains within them were exposed on the asteroid surface to the solar-wind (Bischoff et al., 2006). However, unlike lunar regolith breccias, they contain only a small fraction of grains that resided in the top meter. On asteroids, impacts thoroughly mixed surface material to greater depths before compaction of the regolith breccias (Bischoff et al., 2006).

Most ordinary chondrite regolith breccias are predominantly composed of type 4-6 clasts embedded in a fragmental matrix. Regolith breccias composed predominantly of type 3 material are uncommon. Type 3 clasts are also rare but the matrices of regolith breccias typically contain a few percent of finely dispersed type 3 material including chondrules probably derived from weakly consolidated type 3 chondritic material (e.g., Rubin et al., 1983; Metzler et al., 2011). Thus type 3 material appears to have been mixed by impacts to considerable depths and was not the dominant rock type on the surfaces of the ordinary chondrite parent asteroids when the solar wind gases were acquired. Regolith breccias contain only a few percent or less of impact melted material with textures indicating rapid cooling near the surface, not slow cooling at depth. The abundance of chondrite regolith breccias and their cosmic-ray exposure ages show that near surface material was widely dispersed throughout the parent asteroids (Crabb and Schultz, 1981; Graf and Marti, 1995).

Additional evidence for understanding the role of impacts comes from the metallographic cooling rates of taenite grains in the matrix and in clasts of regolith breccias (Scott and Rajan, 1980; Taylor et al., 1987; Williams et al., 2000). Since the Fe-Ni grains were first etched to ensure they had preserved cloudy taenite and other features of metal in unshocked and unbrecciated chondrites, we can be confident that the cooling rates recorded by these grains were not compromised by impact heating. Figure 11 shows the metallographic cooling rates recorded in the clasts and matrix grains of four H chondrite regolith breccias (Williams et al., 2000). Only Nulles, which is almost entirely composed of type 5 material, has a limited range of clast cooling rates; the others have clasts that cooled at diverse rates of 5-100 °C/Myr. This is the same range that we observe in all but the fast-cooled H4 chondrites. Matrix grains within each chondrite also vary widely in their cooling rates up to 1000 °C/Myr. These metallographic cooling rates confirm that impacts were very effective after metamorphism in mixing the sampled region of the H chondrite parent body. They also show that impact heating was relatively minor.

### 5.5.3. Mezö-Madaras L3 breccia
The L3 chondrite, Mezö-Madaras is a breccia containing L4, carbonaceous, and impact melted clasts. Metal grains in the L3 host and clast have the same metallographic cooling rate of 2 °C/Myr (Scott and Rajan, 1981). The components of the breccia were therefore assembled and consolidated after the L4 material had formed but before the type 3 host material had cooled through 500 °C. This breccia shows that the L chondrite parent body did not have an undisturbed onion-shell structure when it cooled. Similar examples have not been identified among H chondrites but they may exist, e.g, ALH 77299.

### 5.5.4. Portales Valley H6 chondrite
Portales Valley is a unique H chondrite that consists of H6 chondrite clasts separated by metallic Fe-Ni veins (Fig. 12a; Rubin et al., 2001; Ruzicka et al., 2005). Its Ar-Ar age of 4469±12 Myr (Bogard and Garrison, 2009) and metallographic cooling rate of 25 °C/Myr are near the middle of the ranges for H6 chondrites (Figs. 5a and 7). High-precision oxygen isotopic analysis shows that the oxygen isotopic composition of Portales Valley closely matches those of H4-6 chondrites (Folco et al., 2004; McDermott et al., 2012). It is therefore highly probable that it was formed on the H chondrite parent body by impact during metamorphism.

The metal veins in Portales Valley have large continuous Widmanstätten patterns showing that the metal crystallized from the melt very slowly forming single crystals many cm in length. Troilite crystallized subsequently from the residual Fe-Ni-S melt and formed crystals many cm in length that fill the finest veins and permeate the chondritic clasts (Fig. 12b). The chondritic clasts are depleted in metallic Fe-Ni and enriched in FeS although their silicates were not melted (Rubin et al., 2001). These characteristics suggest that the metallic Fe-Ni melt segregated during impact deformation at considerable depth in the H chondrite parent body during formation of pseudotachylite veins by rapid shearing (Tomkins et al., 2013; Melosh, 2005). Portales Valley is shock stage S1 (Table 2), although it could have been shocked to S2 or S3 levels and then annealed causing some recovery (Hutson et al., 2007). It therefore seems likely that the target was near peak metamorphic temperatures prior to the impact so that moderate shock and frictional heating from a major impact could have melted metal-troilite without causing extensive shock damage to the silicates like that seen in S4-6 chondrites.

Although Portales Valley is unique, other H6 S1-S2 chondrites contain evidence for shock mobilization of metal. Metallic veins resembling the narrow millimeter-wide veins in Portales Valley are present in Kernouvé, which has many cm-long metallic veins and one that exceeds 11 cm in length (Christophe Michel-Levy 1981; Hutchison et al., 2001; Friedrich et al., 2013). Butsura has a 30 cm long millimeter-wide vein of metallic Fe-Ni (Hutchison et al., 2001). These veins like those in Portales Valley were probably formed during metamorphism.

In summary, the lack of slowly cooled impact melts in regolith breccias and meteorites and the near absence of slowly cooled meteorites with Ar-Ar ages under 4 Gyr suggest that deeply buried impact melts, as modeled by Davison et al. (2012), did not cause significant metamorphism of ordinary chondrites. Thus metamorphosed chondrites are unlikely to have been reheated by impact above ~700°C so that they cooled slowly enough for cloudy taenite to reform. We therefore infer that chondrites with cloudy taenite have preserved their early thermal histories and that their metallographic cooling rates were probably not modified or reset by impact heating.

Portales Valley and other low-shock H6 chondrites provide direct evidence for localized impact-driven melting and mobilization of Fe-Ni-S at depth during metamorphism. In addition, the L chondrite breccia, Mezö-Madaras, shows that impacts fragmented and mixed material including projectile material during chondrite metamorphism. Regolith breccias provide extensive evidence for thorough impact mixing in the ordinary chondrite parent bodies during the 4 Gyr after metamorphism.

## 5.7. Early impact history of H chondrites

In addition to the macroscopic evidence for impact processing near peak metamorphism described above, there is evidence from microscopic features and Ar-Ar ages for impacts during H chondrite metamorphism. Shock stage S1-S2 chondrites contain microscopic evidence for localized heating and melting of phases. Ashworth (1981) documented TEM evidence for heterogeneous shock effects, for example, extremely localized intense deformation in Butsura (H6 S2). Numerical modeling by Bland et al. (2012) showed that low shock pressures cause large localized temperature excursions in highly porous targets. Rubin (2004) identified numerous microscopic features in thin sections of S1 and S2 chondrites that may have resulted from this process. For example, clusters of irregularly shaped troilite grains can be found in some taenite grains next to larger grains of troilite. These probably formed by highly localized shock melting during peak metamorphism as the sulfides share a common optical orientation and are rimmed by tetrataenite (Scott et al., 2010). The ubiquitous nature of this feature suggests that impact-driven shear processes and mild to moderate shock caused extensive localized melting of troilite in shock stage S1-S2 ordinary chondrites during metamorphism.

Ar-Ar ages of the oldest heavily shocked or impact–melted H chondrites provide direct evidence for impacts during the final stages of metamorphism. An impact clast in the H6 regolith breccia Ourique has an age of 4360±120 Myr and the shock stage S6 H6 chondrite, Y-75100, contains impact melt veins that formed 4490±70 Myr ago (Swindle et al., 2009, 2013). These ages overlap the Ar-Ar ages of slowly cooled H chondrites (Fig. 7). [Korochantseva et al. (2008) claim that the H5 chondrite, Dhofar 323, contains an impact melt clast with an Ar-Ar age of 4508±9 Myr. However, the evidence for impact melting is not clear-cut and Schwarz et al. (2006) infer that the data support the onion-shell model.] Two shock-melted L chondrites also have ancient Ar-Ar ages: Shaw and MIL 05029, have respective ages of 4415±30 and 4517±11 Myr (Turner et al., 1978; Weirich et al., 2010).

We infer that the geological history of ordinary chondrites cannot be simply divided into an early metamorphic era that lasted 150 Myr followed by 4 Gyr of impact modification. Impacts were important from the start. Even though these impacts failed to cause global heating, they must have fragmented and mixed material to great depths. Clearly the H chondrite parent body, like the L parent body (Taylor et al., 1987; Göpel et al., 1994), did not cool from peak metamorphic temperatures with an undisturbed onion shell structure.

## 5.8. Thermal and impact history of H chondrite parent body during metamorphism

Impacts were particularly important during the first ~30 Myr of metamorphism of the H chondrite parent body. The three fast cooled H4 chondrites were excavated by one or more impacts from depths of several km ~3-4 Myr after the H parent body accreted when temperatures would have been close to peak values. [For temperature-time plots as a function of depth see Kleine et al. (2008) or Ganguly et al. (2013).] Evidence from Portales Valley and other H6 chondrites with metallic veins requires at least one major impact in the first ~30 Myr before temperatures had dropped by >100 °C. Metallographic cooling rates and cloudy taenite data show that some H6 chondrites cooled faster below 500 °C than some H4 chondrites indicating that petrologic types 4 and 6 were mixed by impacts during metamorphism. The Ar-Ar ages of H4-6 chondrites also require mixing of petrologic types before the Ar closure temperature of ~300 °C was reached. Since Ar-Ar ages are inversely correlated with cloudy taenite particle size (Fig. 8), cooling rates at ~300 °C may have been correlated inversely with depth. In this case, thermal conduction could have reestablished a radial thermal gradient in the H chondrite body prior to cooling through ~300 °C. The rapid cooling above 700 °C of four H5-6 chondrites at 25-100 °C/kyr inferred by Ganguly et al. (2013) suggests that the sampled portion of the H chondrite parent body was fragmented into km-sized pieces and reaccreted before it cooled to 700°C.

Other meteorite evidence points to the importance of major impacts in the early solar system. Impact melts formed on the parent bodies of enstatite chondrites and achondrites as early as ~4 Myr after CAIs (Keil et al., 1989; Bogard et al., 2010). Group IAB irons with angular silicate clasts, which have textures like those in Portales Valley (Rubin et al., 2001; Ruzicka et al., 2005; Tomkins et al., 2013), were formed within the first 15 Myr of solar system history (e.g., Theis et al., 2013). The parent body of the IIE irons, which resembled the H chondrite body, suffered several major impacts in the first 30 Myr (Schulz et al., 2012). The IVA iron meteorite parent body was destroyed so that the core could cool rapidly within 2-3 Myr after CAI formation (Yang et al., 2008; Blichert-Toft et al., 2010). Given that the mass of the asteroid belt was initially many times its current mass, impact rates would inevitably have been much higher during the time of peak metamorphism than during subsequent slow cooling (Davison et al., 2013; Marchi et al., 2013).

The impact on the H chondrite parent asteroid 3-4 Myr after it accreted may have been a hypervelocity impact that excavated to depths of several km (e.g., Ciesla et al., 2013), or it may have been a grazing impact at near-escape velocity when planetesimals were still accreting (Asphaug et al., 2006). Similarly, the formation of the Portales Valley breccia during the first ~30 Myr may have resulted from a hypervelocity impact that created a crater at least 20 km in diameter (Kring et al., 1999). Alternatively, it may have formed from shear and frictional heating from a glancing low speed impact. A hypervelocity impact would have caused early whole-rock melting nearer the surface for which we have no direct evidence. But given the small number of dated impact melts and the dominant effects of impacts at unusually high velocities (Marchi et al., 2013), we cannot exclude an early hypervelocity impact.

Taylor et al. (1987) inferred that chondrites that cooled at 10 K/Myr were buried at depths of ~40 km and argued that asteroids excavated to this depth would have been catastrophically disrupted and reassembled. However, our thermal model (Fig. 10) suggests that chondrites that cooled at 10 K/Myr may have been excavated from depths of ~15 km by impact craters. Our metallographic cooling rates, excluding those for the fast-cooled H4s, are not correlated with petrologic type, as Taylor et al. (1987) also observed. If the weak correlation between petrologic type and cloudy taenite particle size (Fig. 2) is significant, the H chondrite parent body may once have had an onion-shell structure that was modified by impact cratering during cooling, as Harrison and Grimm (2010) suggested.

To test our conclusions and fully understand the earliest thermal and impact history of the H chondrite parent asteroid, many more exceptionally well-preserved chondrites should be studied by diverse techniques. Chondrites that have been affected by impacts (e.g., Portales Valley) should not be excluded. The sole criterion should be an old Ar-Ar or Pb-Pb age and lack of terrestrial weathering. Nearly all ancient H chondrites are low shock or unshocked (a few are shock stage S3), they have slow cooling rates at 500°C, and possess cloudy taenite rims. However, these are not primary requirements.

## 6. CONCLUSIONS

Our combined studies of cloudy taenite, metallographic cooling rates, and shock levels in H3-6 chondrites show that metallographic cooling rates were not compromised by shock reheating and that the parent asteroid did not cool with an undisturbed onion shell structure. The dimensions of the tetrataenite particles in the cloudy zone correlate inversely with cooling rate as they do in iron and stony iron meteorites confirming that cooling rates of types 3-6 overlapped considerably. Some type 4 chondrites cooled slower than some type 6 chondrites, and type 3 chondrites did not cool faster than types 4-6.

The H4 chondrites that Trieloff et al. (2003) used to validate the onion-shell model cooled through 500 °C at >5000 °C/Myr, at least 50× faster than any sample in a body that was heated by [26]Al and cooled without impact disturbance. Since such fast-cooled H4 chondrites are rare and cooling rates of types 3-6 overlap considerably, we infer that the suite of chondrites studied by Trieloff et al. (2003) are simply not representative of unreheated H4-6 chondrites.

Metamorphism by deep burial of impact-melts (Davison et al., 2012) is not consistent with the absence of slowly cooled impact melts. Impact melted H chondrites that formed during the late heavy bombardment cooled rapidly in hours or less. Regolith breccias show that post-metamorphic impacts fragmented and thoroughly mixed all petrologic types without retention of more than a few vol. % of impact melts. Taenite grains in clasts in regolith breccias commonly preserved their metamorphic history against erasure by impact heating. These observations favor [26]Al as the major heat source for metamorphism.

The three fast-cooled H4 chondrites, Ste. Marguerite, Beaver Creek, and Forest Vale, cooled through 500 °C at ≥5000 °C/Myr. They were probably excavated by one or more impacts ~3-4 Myr after the H chondrite body accreted and cooled in an ejecta blanket. Portales Valley and metallic veins in other H6 S1 chondrites show that impacts also affected type 6 chondrites near peak metamorphic temperatures causing localized melting of metallic Fe-Ni and troilite due to mild to moderate shock and frictional heating during impact-induced shearing. Impacts mixed materials from different depths before they cooled through 500 °C. If the H chondrite body originally developed an onion-shell structure, it was heavily modified by impacts during cooling.

Published Ar-Ar ages of types 4-6 H chondrites do not decrease systematically with increasing petrologic type as would be expected if the parent body was undisturbed by impacts during metamorphism. Ar-Ar ages correlate inversely with cloudy taenite particle size (Fig. 8) suggesting that impact mixing of materials at diverse depths declined during metamorphism allowing internal thermal equilibration to develop a radial thermal gradient when the interior cooled through 300 °C.

ACKNOWLEDGEMENTS

We thank the curators at the American Museum of Natural History, NASA Johnson Space Center, University of New Mexico, and US National Museum of Natural History who loaned meteorite sections for this study, especially Linda Welzenbach and Kevin Righter. Joseph Michael and Paul Kotula, Sandia National Laboratories, kindly assisted with transmission electron microscopy of the two H4 chondrites. We also thank numerous colleagues for valuable discussions, especially Jeff Taylor, Myriam Telus, Greg Herzog, and Jiba Ganguly, and the reviewers, Tim Swindle, John Wasson, and one who was anonymous, for their comments. Early contributions to this study were made by Mr. Mark Greene, presently a graduate student at University of Michigan, during a NSF sponsored summer program at UMass. This research was supported by the NASA cosmochemistry program through grants NNX12AK68G (ES) and NN08AG53G and NNX11AF62G (JIG). Thermal modeling was partially carried out by SW on the general-purpose PC farm at Center for Computational Astrophysics, National Astronomical Observatory of Japan.

Table 1. Sections of H3-6 chondrites studied for determination of cooling rate, shock level, and cloudy taenite particle size.*

| Chondrite[+] | Type | Cooling rate | Shock | Cloudy taenite |
|---|---|---|---|---|
| ALH 77299 | 3.7 | JSC ,77 | JSC ,77 | JSC ,85 |
| Allegan | 5 | | UH 143 | USNM 215 |
| Ankober | 4 | USNM 3399 | USNM 3399-1 | USNM 3399-1 |
| Avanhandava | 4 | UNM 141 | UNM 141 | |
| Bath | 4 | USNM 201-3 | USNM 201-3 USNM 201 | USNM 201-2 |
| Beaver Creek | 4 | | UNM 606 | UNM 606 |
| Butsura | 6 | USNM 6707-2 | USNM 6707-2 | USNM 6707 |
| Conquista | 4 | | UNM 87 | UNM 199 |
| Dhajala | 3.8 | UNM 300 | UNM 300 | UNM 192 |
| EET 86802 | 4 | JSC ,10 | JSC ,11 | |
| Ehole | 5 | USNM 6719 | | USNM 6719 |
| Estacado | 6 | | UH 118 | USNM no # |
| Forest City | 5 | | | USNM unk |
| Forest Vale | 4 | USNM 2319-3 | USNM 2319-3 | USNM 2319-3 |
| GEO 99101 | 4 | JSC ,16 | JSC ,16 | |
| Guareña | 6 | UH 72 | | USNM 1469 |
| Kernouvé | 6 | USNM 2211-1 | USNM 2211-3 | USNM 1054 |
| Landreth Draw | 5 | USNM 6978-2 | | USNM 6978-2 |
| Marilia | 4 | UNM 130 | UNM 130 | |
| Menow | 4 | USNM 2908-2 | | USNM 6748 |
| Mt. Browne | 6 | | USNM 478-1 | USNM 478 |
| Ochansk | 4 | USNM 1788 | UNM 322 | USNM 1788 |
| OTT 80301 | 3.8 | JSC ,5 ,40 | JSC ,5 | JSC ,40 |
| Portales Valley chondritic | 6 | USNM 6975-1 | USNM 6975-1 | USNM 6975-3 |
| Portales Valley metal | 6 | | | AMNH 4978-5 |
| Queens Mercy | 6 | USNM 3486-1 | USNM 3486-1 | USNM 3486 |
| Quenggouk | 4 | UNM 622 | | USNM 452 |
| Richardton | 5 | | | USNM 595 |
| Sena | 4 | USNM 1473-1 | USNM  1473-1 | USNM 1473 |
| Ste. Marguerite | 4 | USNM 7212-1 | | USNM 7212-2 |
| Tieschitz | H/L3.6 | | | USNM 6783 |

* Sources: AMNH, American Museum of Natural History; JSC, NASA Johnson Space Center; UH, University of Hawaii; UNM, University of New Mexico; USNM, US National Museum of Natural History.
[+] All are falls except for the four Antarctic chondrites (ALH, Allan Hills; EET, Elephant Moraine; GEO, Geologists Range; and OTT, Outpost Nunatak), Landreth Draw, and Estacado.

Table 2. Shock stages for 31 H3-6 chondrites.

| Chondrite | Type | This work[†] | Stöffler et al. (1991) | Stöffler et al. (1992)* | Rubin (2004) | Other# |
|---|---|---|---|---|---|---|
| ALH 77299 | 3.7 | S1-3 br^ | | | | |
| Allegan | 5 | S1 | | | S1 | |
| Ankober | 4 | S1 | | | | |
| Avanhandava | 4 | S1(2) | S2 | | S2 | |
| Bath | 4 | S2 | | | | |
| Beaver Creek | 4 | S1 | | S1 | | S3 |
| Butsura | 6 | S2 | | | | |
| Conquista | 4 | S1 | | S1/S2 | | |
| Dhajala | 3.8 | S1 | S1 | | | |
| EET 86802 | 4 | S1 | | | | |
| Estacado | 6 | S1 | | S1 | S1 | |
| Forest City | 5 | | S2 | | S2 | |
| Forest Vale | 4 | S1 | | | S2 | |
| GEO 99101 | 4 | S1 | | | | |
| Guareña | 6 | | | S1 | S1 | |
| Kernouvé | 6 | S1 | | S1 | S1 | |
| Landreth Draw | 5 | | | | | S2 |
| Marilia | 4 | S3 | S3 | | | |
| Menow | 4 | | | S1 | S2 | |
| Mt. Browne | 6 | S3(2) | | | | S3 |
| Nuevo Mercurio | 6 | S1 | | S1 | | |
| Ochansk | 4 | S3 | S3 | | | S3 |
| OTT 80301 | 3.8 | S2 | | | | |
| Portales Valley | 6 | S1 | | | S1 | |
| Queens Mercy | 6 | S2(3) | | | S2 | |
| Quenggouk | 4 | | | S2(1) | | |
| Richardton | 5 | S1 | | S1(2) | S2 | |
| Sena | 4 | S1 | | | | |
| Ste. Marguerite | 4 | | | | S2 | |
| Tieschitz | 3.6 | | | S1(2) | | |

* Data in Stöffler (private communication, 1992).
[†] Cases where the proportion of grains with the highest shock level was close to the 25% cutoff are indicated by parentheses.
^ Breccia.
# Rubin (1994) except for Landreth Draw, which is from McCoy in Grossman et al. (2000).

Table 3. Metallographic cooling rates and cloudy taenite dimensions for 31 H3-6 chondrites.

| Meteorite | Type | Cooling rate* (°C/Myr) | Cloudy taenite particle size (nm) | ±2 SEM | No. particles |
|---|---|---|---|---|---|
| ALH 77299 | 3.7 | 10 | 90.9 | 2.5 | 178 |
| Dhajala | 3.8 | 100 | 41.9 | 2.0 | 80 |
| OTT 80301 | 3.8 | 15 | 79.4 | 3.4 | 140 |
| Tieschitz | H/L 3.6 | | 82.2 | 3.5 | 120 |
| Ankober | 4 | 7 | 98.5 | 2.6 | 90 |
| Avanhandava | 4 | 20 | | | |
| Bath | 4 | 40 | 64.6 | 4.2 | 90 |
| Beaver Creek | 4 | 5,000 (b) | None | | |
| Conquista | 4 | 25 (b) | 73 | 10 | 90 |
| EET 86802 | 4 | 40 | | | |
| Forest Vale | 4 | 10,000[+] | None | | |
| GEO 99101 | 4 | 30 | | | |
| Marilia | 4 | 6 | | | |
| Menow | 4 | 60[+] | None | | |
| Ochansk | 4 | 15 | 92.4 | 4.8 | 120 |
| Quenggouk | 4 | 30 | 82.9 | 4.0 | 90 |
| Sena | 4 | 20 | 79.8 | 5.7 | 90 |
| Ste. Marguerite | 4 | >10,000 | None | | |
| Allegan | 5 | 15 (b) | 69.8 | 4.1 | 90 |
| Ehole | 5 | 10 | 85.0 | 3.8 | 90 |
| Forest City | 5 | 20 (a) | 87.4 | 6.2 | 90 |
| Landreth Draw | 5 | 20 | 86 | 10 | 57 |
| Richardton | 5 | 20 (b) | 74.2 | 4.8 | 90 |
| Butsura | 6 | 50 | 67.5 | 2.6 | 80 |
| Estacado | 6 | 10 (b) | 124.5 | 12 | 20 |
| Guareña | 6 | 15 | 120.7 | 4.8 | 110 |
| Kernouvé | 6 | 10 | 90.9 | 4.7 | 100 |
| Mt. Browne | 6 | 30 (a) | 67.3 | 3.2 | 70 |
| Portales Valley chondritic | 6 | 25 | 109.0 | 5.2 | 160 |
| Portales Valley metal vein | | | 106.3 | 7.1 | 80 |
| Queens Mercy | 6 | 10 | 82.2 | 4.2 | 90 |

* Cooling rates for eight chondrites that are in italics are from *a*, Willis and Goldstein (1981b) and *b*, Taylor et al (1987).

[+] For Menow and Forest Vale, the precision in the cooling rate is a factor of 4. For the other chondrites, the precision is a factor of 2. The error in the cooling rates is probably a factor of 3-4 (see text).

SEM: standard error of mean.

Table 4. Published metallographic cooling rates and comparison with our study.

| Chondrite | Type | Cooling Rates (°C/Myr) | | | Shock stage | Reference |
|---|---|---|---|---|---|---|
| | | This work | Taylor et al. (1987) | Willis & Goldstein (1981b) | | |
| Bath | H4 | 40 | | 80 | S2 | This work |
| Cee Vee | H5 | | | 25 | | |
| Dhajala | H3 | 100 | 50 | | S1 | This work |
| Ehole | H5 | 10 | | 4 | | |
| Guareña | H6 | 15 | | 6* | S1 | Stöffler et al. (1992) |
| Kernouvé | H6 | 10 | 10 | | S1 | This work |
| Kesen | H4 | | 20 | | S3 | Stöffler et al. (1991) |
| Malotas | H5 | | | 8 | | |
| Nuevo Mercurio | H5 | | 15 | | S1 | This work |
| Queens Mercy | H6 | 10 | 10[a] | | S2 | This work |
| Quenggouk | H4 | 30 | 10[a] | | S2 | Stöffler et al. (1992) |
| Salles | H6 | | | 10 | | |
| Sutton | H5 | | 10[a] | 4 | S2 | Stöffler et al. (1992) |

* Willis & Goldstein (1983) reported 4.3 °C/Myr by averaging all data points. Using the lower envelope of points gives 6 °C/Myr.
[a] Data of lower quality due to scatter on Wood plot.

Table 5. Cooling times inferred from radiometric ages*, metallographic cooling rates, and cloudy taenite particle sizes for eight H4-6 chondrites studied by Trieloff et al. (2003).

| Meteorite | Type | Cooling Times | | Cooling Rate (°C/Myr) | Cloudy taenite (nm) |
| | | Pb-Pb[^] (Myr) | Ar-Ar[^] (Myr) | | |
|---|---|---|---|---|---|
| Ste. Marguerite | H4 | 4 | 5 | >10$^4$ | None |
| Forest Vale | H4 | 6 | 15 | 10$^4$ | None |
| Allegan | H5 | 17 | 26 | 15 | 70 ± 4 |
| Richardton | H5 | 16 | 42 | 20 | 74 ± 5 |
| Sena[+] | *H5*[+] | 11 | - | 20 | 80 ± 6 |
| Kernouvé | H6 | 45 | 68 | 10 | 91 ± 5 |
| Estacado | H6 | - | 102 | 10 | 124 ± 12 |
| Guareña | H6 | 63 | 109[#] | 15 | 121 ± 5 |

* Time interval between CAI formation at 4567 Myr (Connelly et al., 2012) and the Pb and Ar closure temperatures of 720±50 and 550±20 K, respectively (Trieloff et al., 2003).

[^] Ar-Ar ages have uncertainties of 5-16 Myr; Pb-Pb ages have uncertainties of 5-16 Myr. The published Ar-Ar ages were first corrected using an updated $^{40}$K decay constant that adds 30 Myr (see Henke et al., 2012).

[+] Sena is listed here as an H5 following Trieloff et al. (2003). However, Van Schmus and Wood (1967) classified it as an H4 chondrite (see text).

[#] Ar-Ar age from Henke et al. (2012).

Table 6. Radiometric dating and thermal modeling of H4-6 chondrites.

| Meteorite | Type | Turner et al. (1978)[a] | Pellas & Storzer (1981) | Göpel et al. (1994) | Brazzle et al. (1999)[b] | Trieloff et al. (2003) | Kleine et al. (2008)[c] | Henke et al. (2012) | Monnereau et al. (2013) |
|---|---|---|---|---|---|---|---|---|---|
| Beaver Creek | H4 | | x | | x | | | | |
| Forest Vale | H4 | x | | x | | x | | x | x |
| Ste. Marguerite | H4 | | | x | x | x | x | x | x |
| Sena | *H4*[d] | | | | | x | | | |
| Allegan | H5 | | x | x | x | x | | x | x |
| Nadiabondi | H5 | | x | x | | x | | x | x |
| Richardton | H5 | x | | x | x | x | x | x | x |
| Estacado | H6 | | x | | | x | x | x | x x |
| Guareña | H6 | x | x | x | x | x | | x | |
| Kernouvé | H6 | x | x | x | x | x | x | x | x |

[a] Also studied Tieschitz, Menow, Ochansk, Butsura, Queens Mercy, and Mt. Browne.
[b] Also studied Phum Sambo (H4).
[c] Also studied ALH 84069 (H5).
[d] Note that Trieloff et al. (2003) classified Sena as H5 although Van Schmus and Wood (1967) list it as an H4 chondrite.

Figure Captions

Fig. 1. Polished and etched section of the Allegan H5 chondrite. a. Reflected light image showing taenite grain (white) with a brown etched rim of cloudy taenite (CT) and an outermost unetched rim of tetrataenite (Tt, white), which is 1-3 μm wide. The dark matrix is silicate. b. Scanning electron image of the cloudy taenite rim showing an intergrowth of rounded tetrataenite particles (lighter gray) surrounded by a low-Ni phase (darker gray). The size of the tetrataenite particles is inversely related to the cooling rate. The dark band on the right side is the tetrataenite rim that encloses the taenite grain.

Fig. 2. *a*. Cloudy taenite particle size, which reflects the cooling rate at 300 °C, vs. petrologic type for 21 H3-6 chondrites. Particle sizes in types 3 to 6 show considerable overlap, contrary to the predictions of onion-shell body. Data and meteorite names are given in Table 3. *b*. Symbols show shock stage of chondrites (Table 2). Most are shock stage S1; the few shock stage S2 and S3 chondrites are not systematically displaced. Error bars are 2 standard errors of mean.

Fig. 3. Ni concentration profiles across taenite grains in a) Guareña showing a typical M-shaped Ni profile from silicate on the left to kamacite on the right, and b) Ste. Marguerite. The Ni zoning profile in (a) is symmetrical because of rapid diffusion along metal-silicate boundaries (see Willis and Goldstein, 1983). Taenite grains in Guareña, which cooled at 15 °C/Myr, contain 33-40% Ni except for the outermost tetrataenite rim that contains 45-50% Ni. In Ste. Marguerite taenite grains have uniformly low Ni concentrations of ~10% except for the outermost few micrometers that are enriched in Ni indicating exceptionally rapid cooling at >$10^4$ °C/Myr. (The Ni content on the left side of the taenite in (b) is higher because of adjacent troilite.)

Fig. 4. Ni concentration at the center of taenite grains vs. apparent distance to the nearest edge for a) Dhajala H3 and Kernouvé H6, and b) Ankober H4 and Butsura H6. Calculated cooling rate curves are from Willis and Goldstein (1981b) except for the 1000 °C/Myr curve, which was extrapolated by Taylor et al. (1987). Because grains are sectioned randomly, the lower envelope of data points is used to infer the cooling rate. In (a), Dhajala H3 cooled at 100 °C/Myr and Kernouvé H6 at 10 °C/Myr. By contrast in (b), the lower petrologic type chondrite, Ankober (H4), cooled at 7 °C/Myr, much slower than the H6 chondrite Butsura, which cooled at 50 °C/Myr.

Fig. 5. Metallographic cooling rate vs. petrologic type for 35 H3-6 chondrites: a) shows sources of data; b) shows shock stage of chondrites (Tables 2 and 3). Cooling rate ranges in types 3-6 overlap considerably contrary to onion-shell model and are not correlated with shock stage. Most cooling rates have a precision of a factor of 2.

Fig. 6. Cloudy taenite particle size vs. metallographic cooling rate for 21 H3-6 chondrites. Solid straight line shows the least-squares line for H3-6 chondrites; dashed line is the least-squares line for iron and stony-iron meteorites (Goldstein et al., 2013). The inverse correlation shown by the H chondrite data and the proximity of the least squares lines for chondrites and metal-rich meteorites validates the metallographic cooling rates for the H chondrites. If H3-6 chondrites had cooled in a body with an undisturbed onion-shell structure, type 6 would plot at top left and type 3 at lower right, with types 4 and 5 located between them.

Fig. 7. $^{40}$Ar-$^{39}$Ar ages for 17 type 4-6 H chondrites. Data from Trieloff et al. (2003) show an inverse correlation between age and petrologic type consistent with an onion-shell parent body, but data from Turner et al. (1978), Bogard et al. (2001), Bogard and Garrison (2009), and Schwarz et al. (2006) do not follow the Trieloff et al. trend. Meteorite names: Monahans (1998), Dh: Dhofar 323; others are given in Tables 1 and 6.

Fig. 8. Cloudy taenite particle size vs. $^{40}$Ar-$^{39}$Ar age for 12 H chondrites. Cloudy taenite particle size, which reflects the cooling rate at 300 °C and $^{40}$Ar-$^{39}$Ar age, which dates the time of cooling through this temperature, are inversely correlated, even if the two H4 chondrites, Forest Vale and Ste. Marguerite, that cooled too rapidly to develop cloudy taenite, are omitted. However, more data are needed to test the correlation. Particle sizes are from Table 3; sources of age data are given in Fig. 7 caption.

Fig. 9. Cooling rate at 500 °C vs. depth below the surface calculated for chondritic bodies with radii of 50 and 100 km that accreted 2.0 Myr after CAIs, were heated by $^{26}$Al decay, and cooled without being impacted. Material 4 km below the surface would have had the highest cooling rate of 80 °C/Myr. Cooling rates plummet to zero at a depth of 3 km where the peak temperature was 500 °C. Cooling rates above 100 °C/Myr could not have been achieved in a body that was metamorphosed by $^{26}$Al and cooled undisturbed.

Fig. 10. Cooling rate at 500 °C vs. peak temperature calculated for 100 km radius chondritic body that accreted 2.0 Myr after CAIs and was heated by $^{26}$Al decay. Burial depths are marked along the curve. Vertical dotted lines show approximate boundary temperatures for types 3, 4, 5, and 6 chondrites (Harrison and Grimm, 2010).

Fig. 11. Histogram showing metallographic cooling rates for clasts and ranges of cooling rates for matrix grains in four H chondrite regolith breccias (data from Williams et al., 2000). Cooling rates of the clasts range from 5 to 100 °C/Myr, which is the same range found in all but the three fast cooled H4 chondrites (Fig. 5). Matrix grains tend to show a slightly wider range of cooling rates. Most regolith breccias contain material from nearly all the locations sampled by the unbrecciated samples.

Fig. 12. *a.* The Portales Valley H6 chondrite showing type 6 chondrite clasts separated by metallic Fe-Ni veins (white). *b.* A closer view of the lower left corner where troilite fills the finest veins and permeates the chondrite clasts, which are depleted in metallic Fe,Ni relative to normal H6 chondrites. The Ar-Ar age of Portales Valley of 4469±12 Myr (Bogard and Garrison, 2009), slow cooling rate of 25 °C/Myr (Table 3), and low shock-stage (S1) suggest that the impact that created this breccia occurred during metamorphism of the H chondrite parent body, probably when the interior was close to peak temperatures. Specimen USNM 6975; photographs courtesy Linda Welzenbach.

Fig. 1a, b

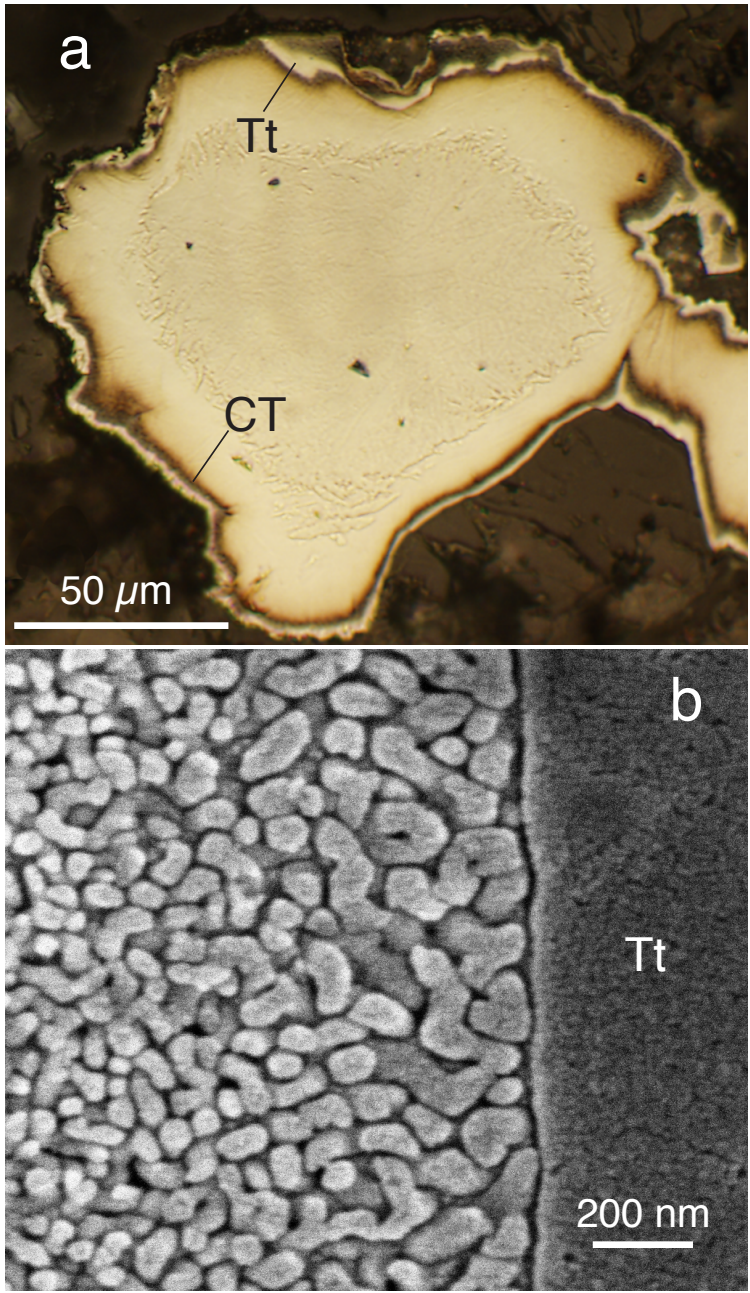

Fig. 2a, b

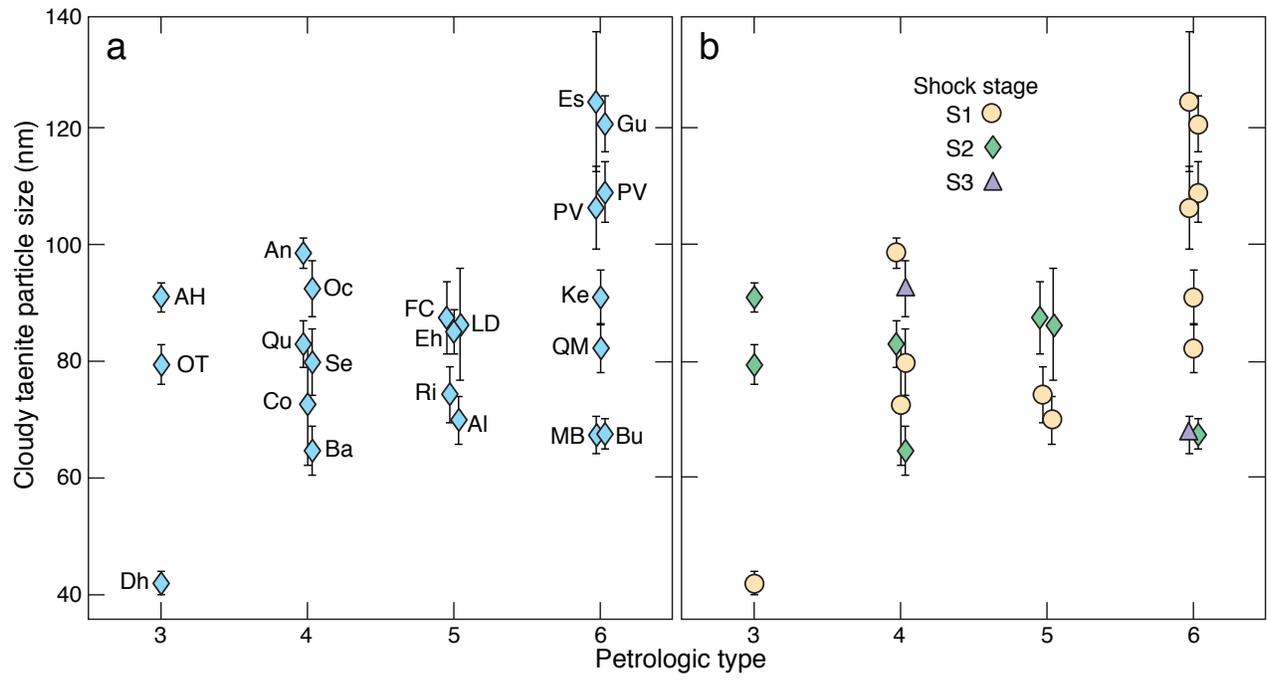

Fig. 3a, b

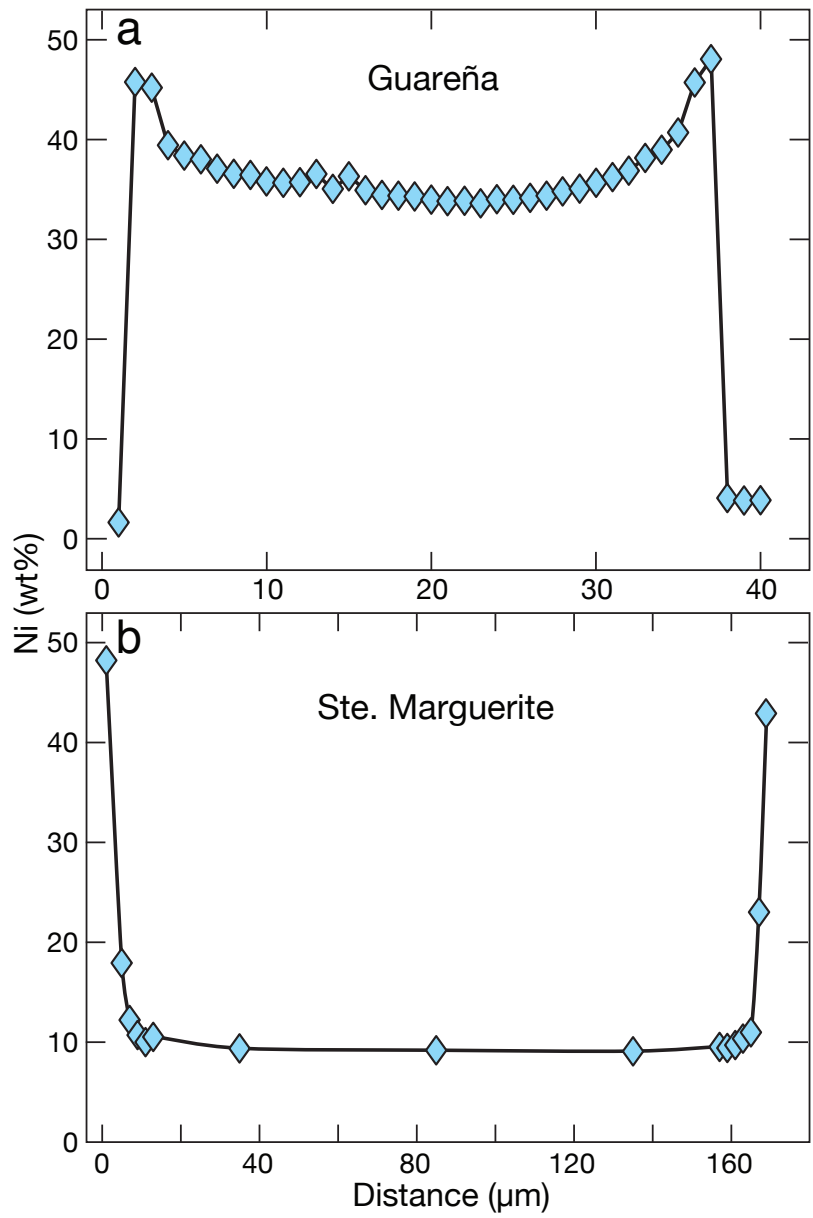

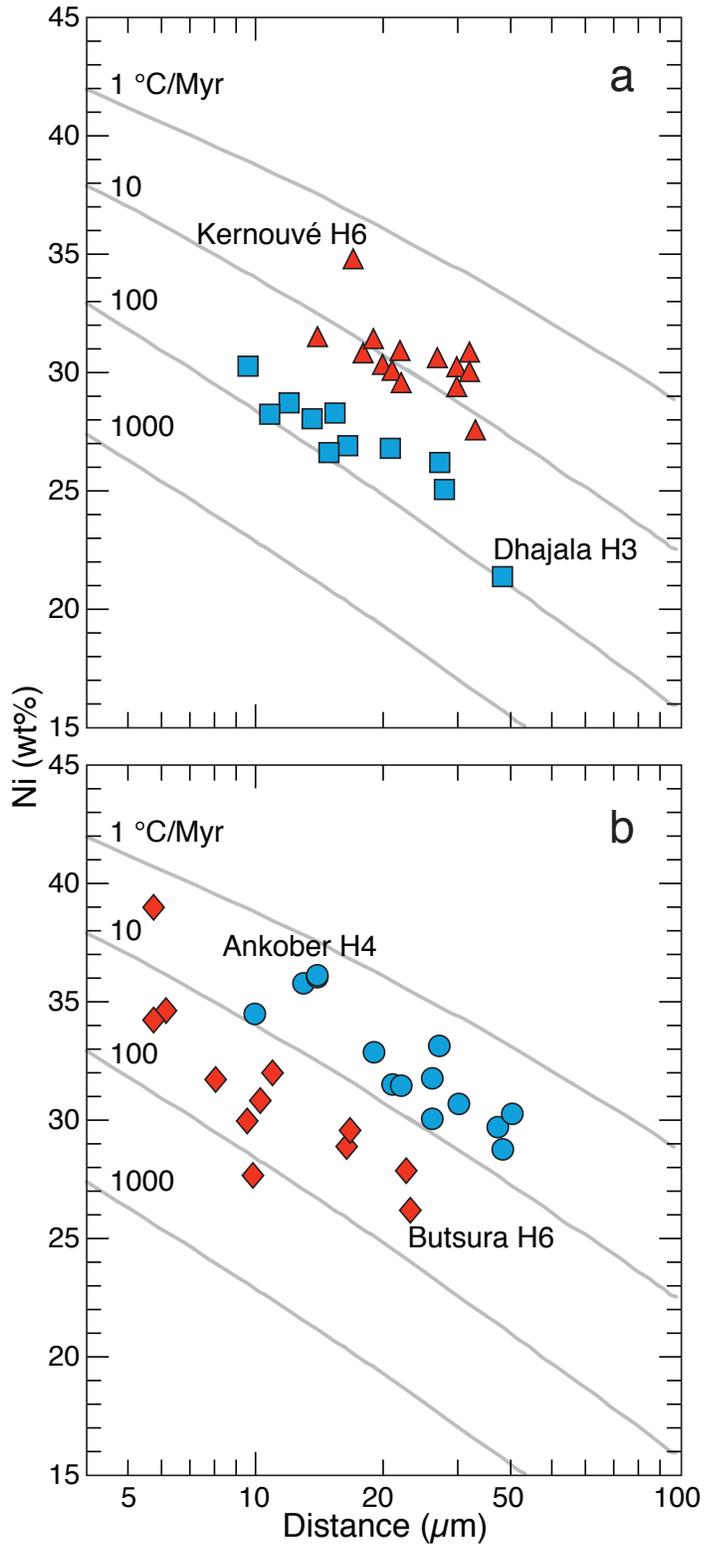

Fig. 5a, b

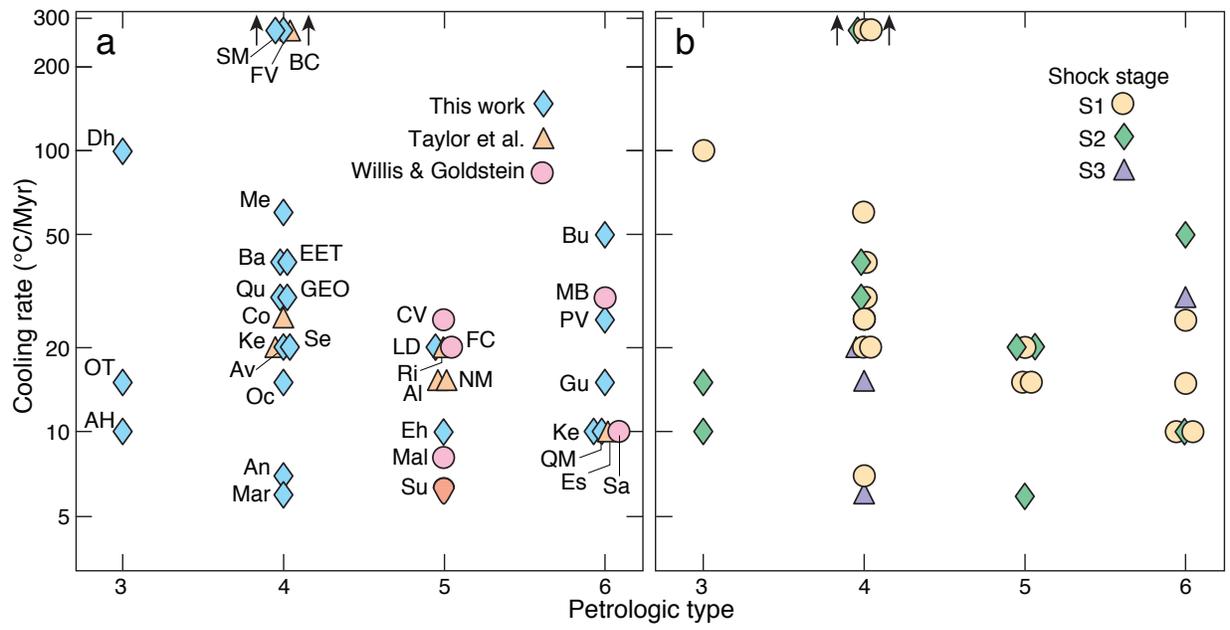

Fig. 6

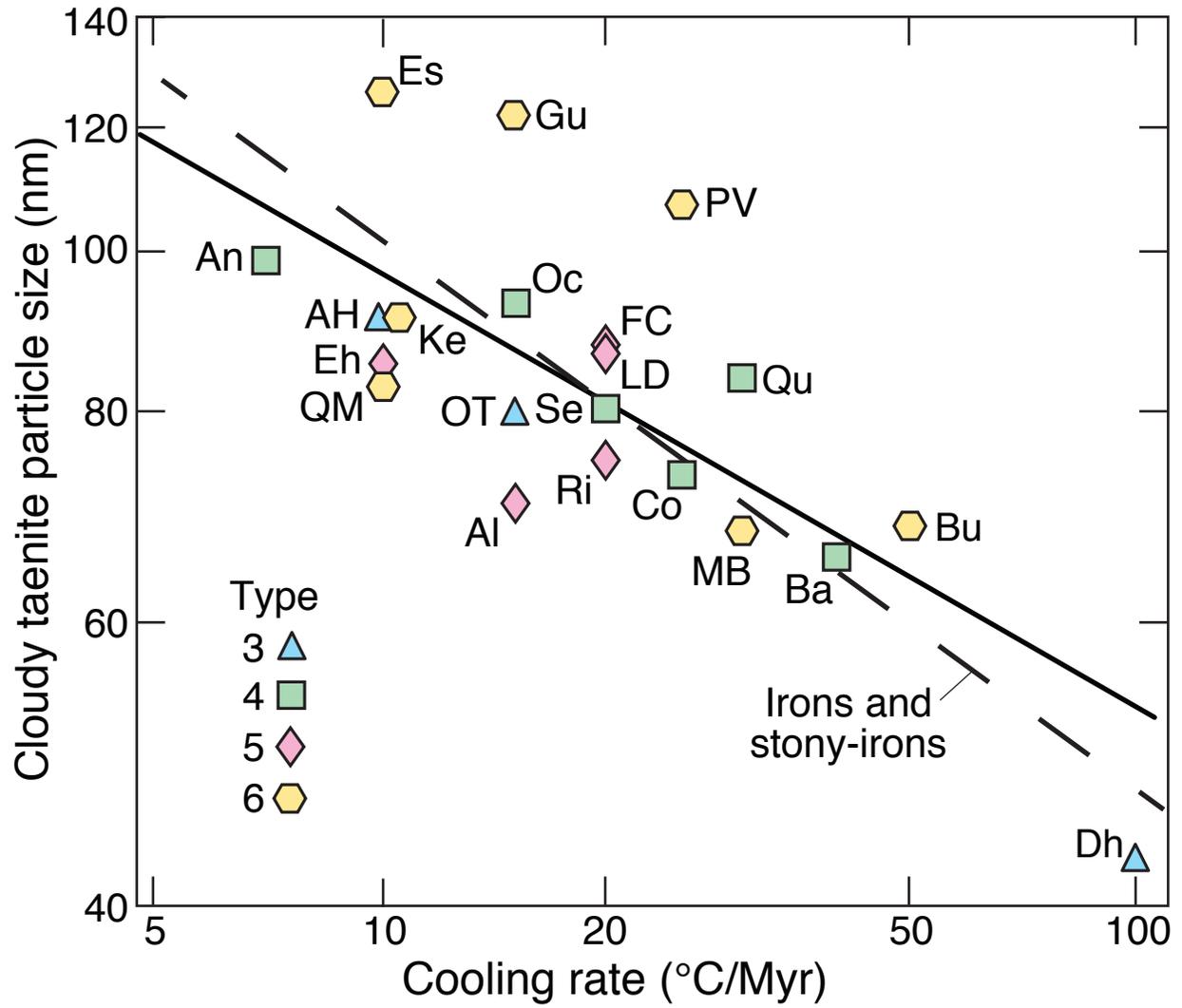

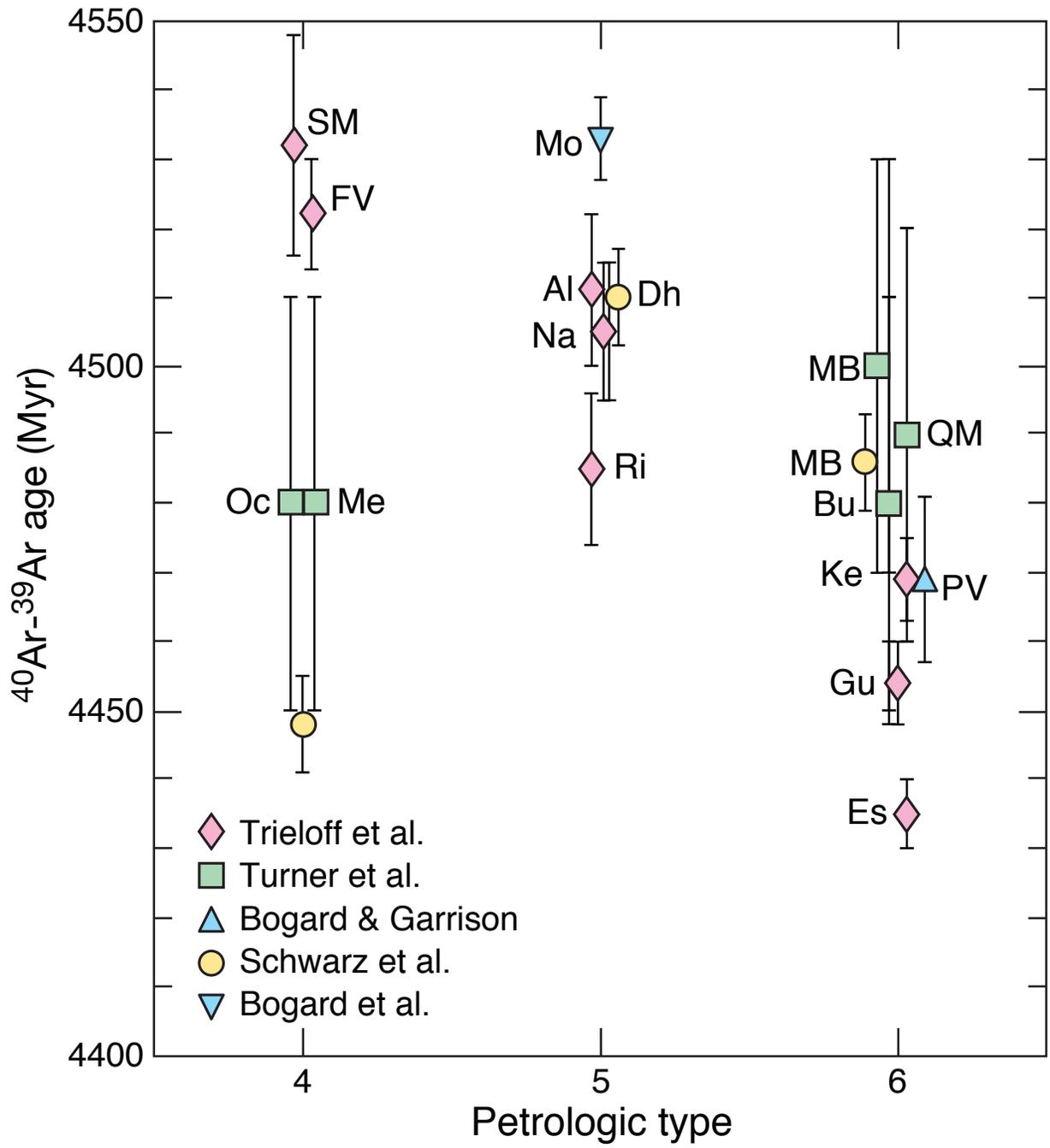

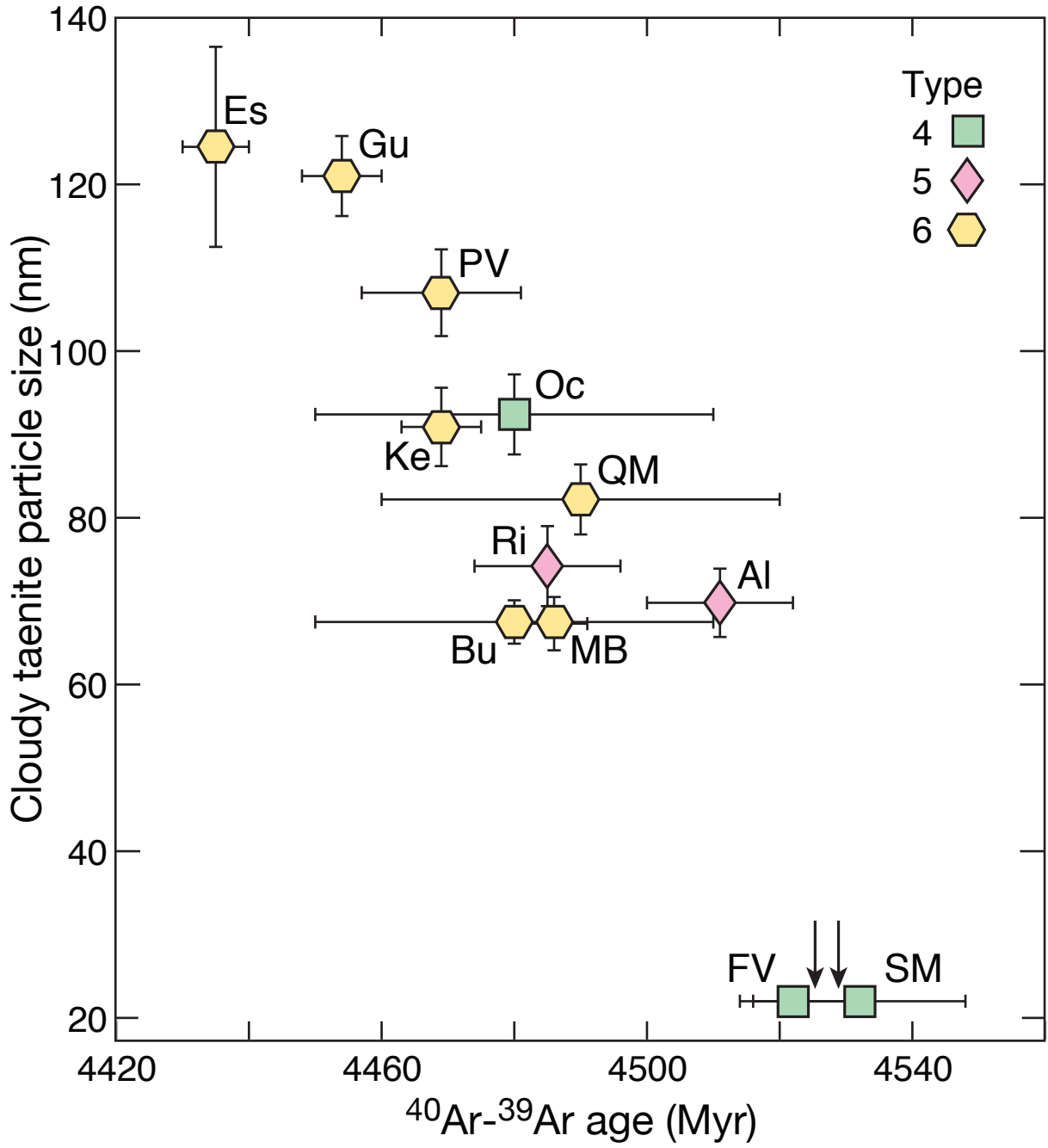

Fig. 9

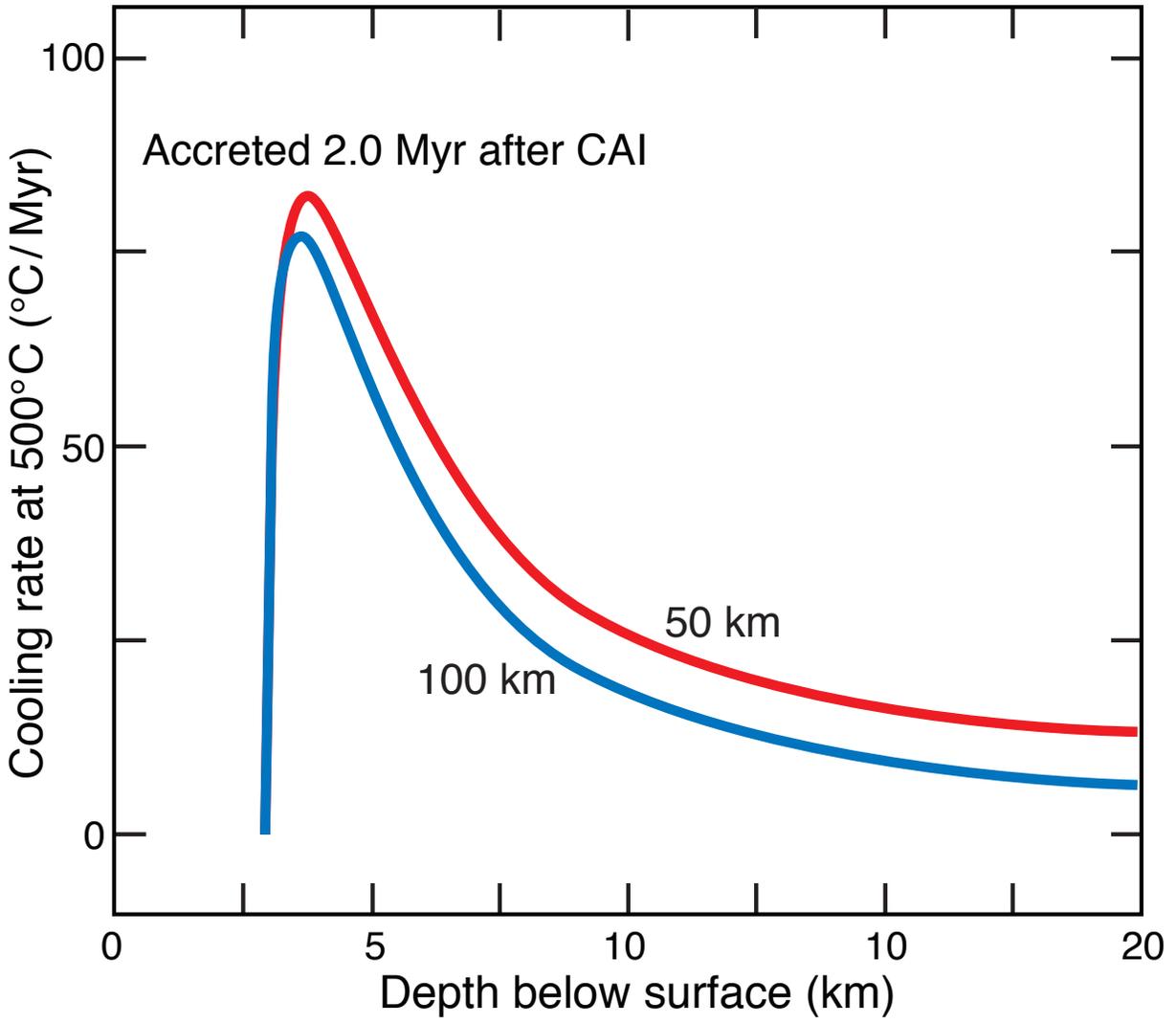

Fig. 10

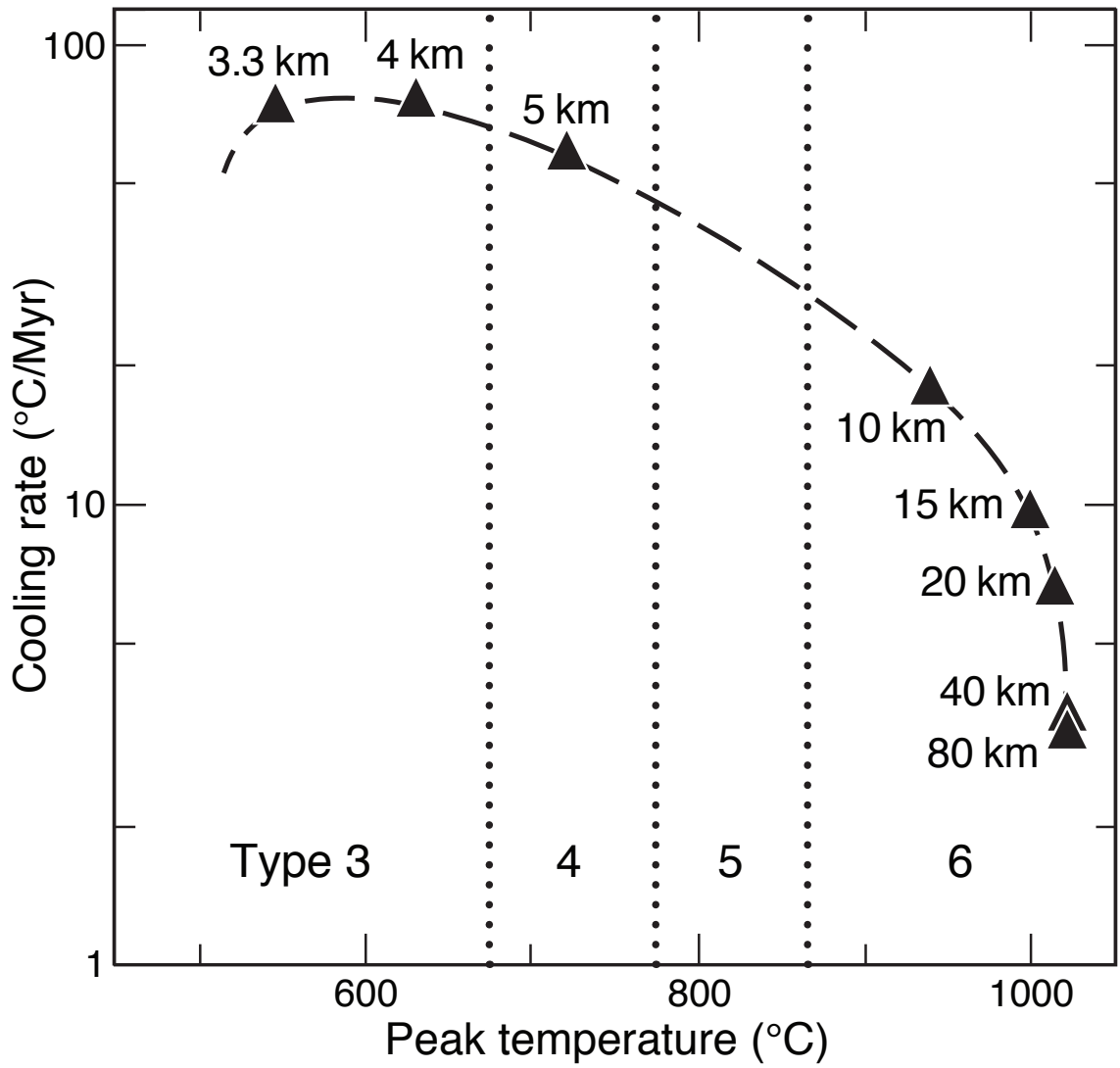

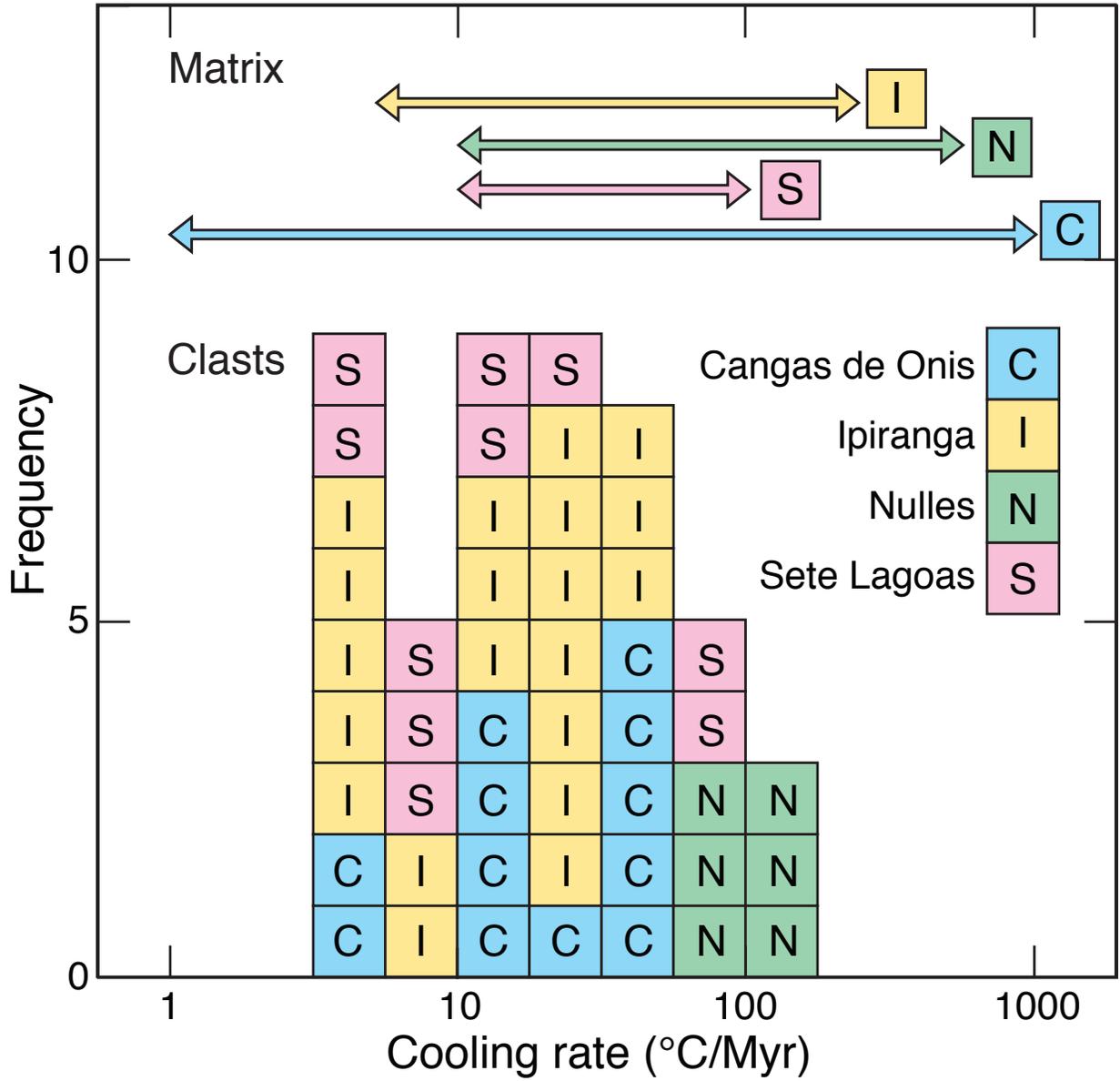

Fig. 12

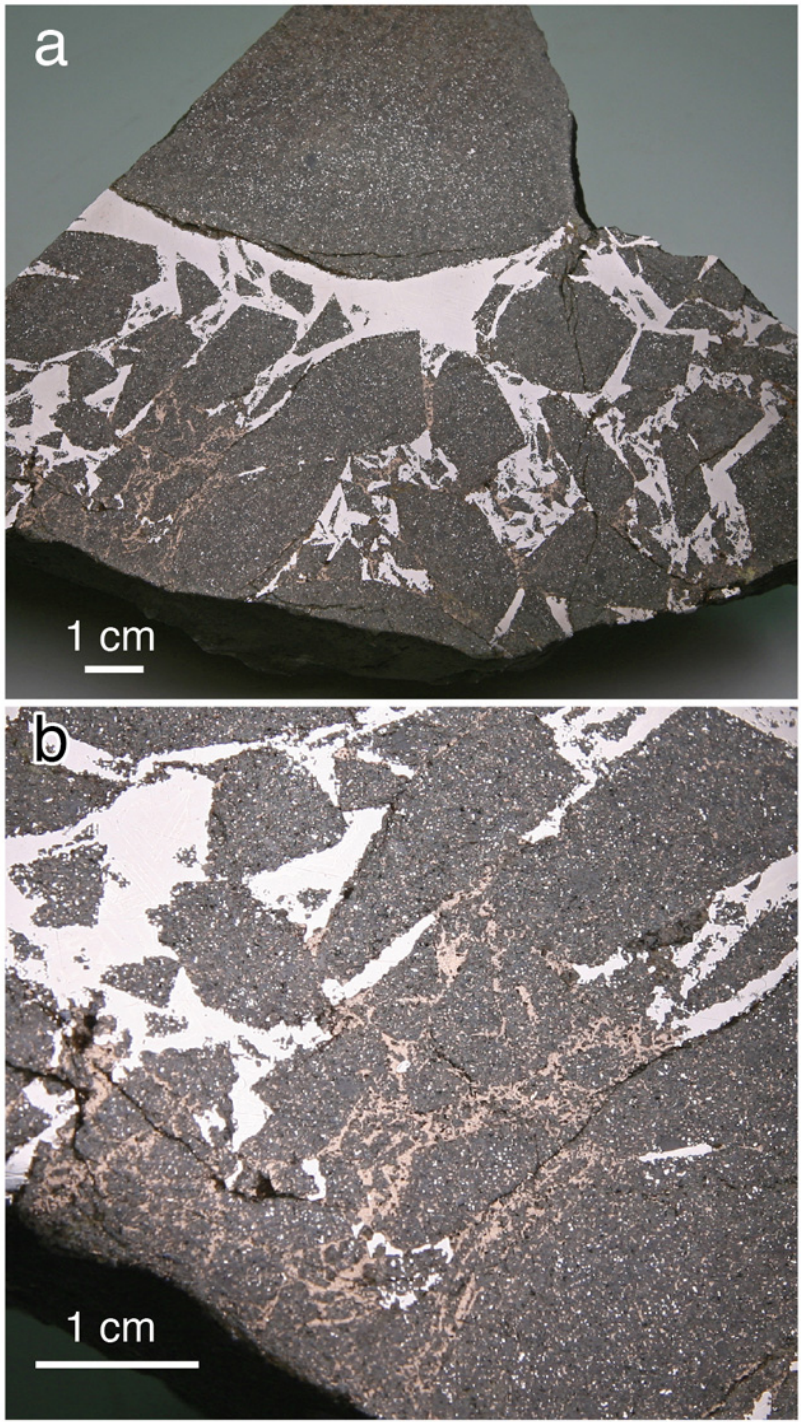